\documentclass{ieeeaccess}
\usepackage{cite}
\usepackage{amsmath,amssymb,amsfonts}
\usepackage{algorithmic}
\usepackage{graphicx}
\usepackage{textcomp}
\usepackage{url}
\usepackage{booktabs, threeparttable, multirow}
\usepackage{xcolor}
\usepackage[normalem]{ulem}

\newcommand\hl[1]{%
  \bgroup
  \hskip0pt\color{blue!80!black}%
  #1%
  \egroup
}

\newcommand\hlr[1]{%
  \bgroup
  \hskip0pt\color{red!80!black}%
  #1%
  \egroup
}

\begin{document}
\history{Date of publication xxxx 00, 0000, date of current version xxxx 00, 0000.}
\doi{10.1109/ACCESS.2017.DOI}

\title{Embedding-based Music Emotion Recognition Using Composite Loss}
\author{
\uppercase{Naoki Takashima}\authorrefmark{1,4},
\uppercase{Frédéric Li}\authorrefmark{2},
\uppercase{Marcin Grzegorzek\authorrefmark{2,5}, and Kimiaki Shirahama}.\authorrefmark{3,6}
}
\address[1]{Graduate School of Science and Engineering, Kindai University, 3-4-1, Kowakae, Higashiosaka, Osaka 577-8502, Japan (e-mail: naoki.takashima.kindai@gmail.com)}
\address[2]{Institute of Medical Informatics, University of Lübeck, Ratzeburger Allee 160, 23538 Lübeck, Germany (e-mail:  fr.li@uni-luebeck.de, marcin.grzegorzek@uni-luebeck.de)}
\address[3]{Faculty of Informatics, Kindai University, 3-4-1 Kowakae, Higashiosaka, Osaka 577-8502, Japan (e-mail: shirahama@info.kindai.ac.jp)}
\address[4]{Speee Inc., Roppongi Grand Tower FL35, 3-2-1 Roppongi, Minato-ku, Tokyo 106-0032, Japan}
\address[5]{Department of Knowledge Engineering, University of Economics in Katowice, Bogucicka 3, 40287 Katowice, Poland}
\address[6]{Department of Information Systems Design, Doshisha University, 1-3 Tatara Miyakodani, Kyotanabe, Kyoto 610-0394, Japan (e-mail: kshiraha@mail.doshisha.ac.jp)}

\tfootnote{This work has been supported in part by Japan Society for the Promotion of Science (JSPS) within Grant-in-Aid for Scientific Research (B) (19H04172).}

\markboth
{Author \headeretal: Preparation of Papers for IEEE TRANSACTIONS and JOURNALS}
{Author \headeretal: Preparation of Papers for IEEE TRANSACTIONS and JOURNALS}

\corresp{Corresponding author: Kimiaki Shirahama (e-mail: kshiraha@mail.doshisha.ac.jp).}
%.

\begin{abstract}
Most music emotion recognition approaches perform classification or regression that estimates a general emotional category from a distribution of music samples, but without considering emotional variations (e.g., happiness can be further categorised into much, moderate or little happiness). We propose an embedding-based music emotion recognition approach that associates music samples with emotions in a common embedding space by considering both general emotional categories and fine-grained discrimination within each category. Since the association of music samples with emotions is uncertain due to subjective human perceptions, we compute composite loss-based embeddings obtained to maximise two statistical characteristics, one being the correlation between music samples and emotions based on canonical correlation analysis, and the other being a probabilistic similarity between a music sample and an emotion with KL-divergence. The experiments on two benchmark datasets demonstrate the effectiveness of our embedding-based approach, the composite loss and learned acoustic features. In addition, detailed analysis shows that our approach can accomplish robust bidirectional music emotion recognition that not only identifies music samples matching with a specific emotion but also detects emotions expressed in a certain music sample.
\end{abstract}

\begin{keywords}
Music emotion recognition, Embeddings, Canonical correlation analysis, Kullback-Leibler divergence, Bidirectional retrieval
\end{keywords}

\titlepgskip=-15pt

\maketitle

\section{Introduction}
\label{sec:intro}

\PARstart{M}{usic} is a powerful means to evoke human emotions. Analysing the interactions between them is thus important in affective computing, and is one of main focuses of \textit{Music Emotion Recognition} (MER) which attempts to automatically identify the emotion matching a specific music~\cite{han2022_survey}. MER is useful for many potential applications such as music recommendation and playlist generation for streaming services, and even music therapy in biomedicine\cite{Hizlisoy_DL_class}. 

MER is performed differently in the literature depending on several factors. The first one is how emotions are modelised. Two main frameworks to define emotions currently co-exist: the \textit{categorical} and the \textit{continuous} ones. The former defines emotions as explicit categories, either directly using the six `basic' emotions highlighted in Ekman's theory (i.e., happiness, sadness, anger, fear, disgust and surprise)~\cite{Ekman_BasicEmotions} or derivatives from them. The latter decomposes emotions along several axes, among which the most popular axes are arousal (level of energy) and valence (level of pleasantness) based on Russell's Circumplex model~\cite{CircularModel_Russell}. Both of the categorical and continuous frameworks have their pros and cons. While the former can clearly identify general emotions in music, it is not the most appropriate to take into account the richness and variations of human emotions. For example, there are several degrees of happiness ranging from little to intense happiness, that cannot be distinguished from each other with the categorical models. On the other hand, the continuous approach can express fine-grained human emotions in a vector space defined by the arousal and valence axes. However, it is difficult to identify general emotions because dissimilar emotions such as `fear' and `anger' are located close to each other in the arousal-valence space~\cite{CircularModel_Dufour}. Therefore, neither categorical nor continuous approach has become predominant over the other in the literature, despite the benefits of each approach being essential for MER.

The second main difference among MER work lies in how MER is translated into a machine learning problem. The most popular approaches so far have consisted in considering MER either as a \textit{classification} or a \textit{regression} problem, depending on whether the categorical or continuous modelisation of emotions is used~\cite{gomez2021}. Classification and regression methods are however unidirectional, and mostly investigate \textit{Music to Emotion} (M2E). In this framework, a classification or regression model is trained to respectively output categorical emotion estimations or arousal-valence intensity scores given some music-related input data (e.g., audio records, lyrics transcripts, playlist information, etc.). On the other hand, \textit{Emotion to Music} (E2M) aims to retrieve some relevant music extract given some emotion-related input remains more marginal. More recently, \textit{embedding-based retrieval} approaches have emerged to address this issue. In a retrieval problem, a model is trained to return a list of examples ordered in a descending order of similarities to an input example, also referred to as a \textit{query}~\cite{chen2022}. For M2E, the query and retrieved examples are respectively music-related data and emotion, while the reverse is true for E2M. Embedding-based methods on the other hand aim to project examples from various modalities into a common space referred to as an \textit{embedding space}, so that the \textit{embeddings} (i.e. vectorial projections) of two data examples associated to the same concept are close to each other~\cite{chun2021}. The combination of embedding and retrieval approaches enables \textit{bidirectional} retrieval, allowing retrieval methods to perform either E2M or M2E in the case of MER, and consequently has led to an increased interest from the research community over the past years~\cite{Zeng_closs_S-DCCA_audio-image,Zeng_closs_TNN-C-CCA_audio-image,Yu_closs_DCCA_audio-text,won2021-text,shang2021,ferraro2021}. But while audio, image and text modalities are the most popular in the literature to the best of our knowledge, no other work has so far attempted to investigate bidirectional retrieval between audio and emotion ratings, except for our previous study~\cite{CMR_takashima}.

\begin{figure}[tbp]
	\centering 
	\includegraphics[width=\linewidth]{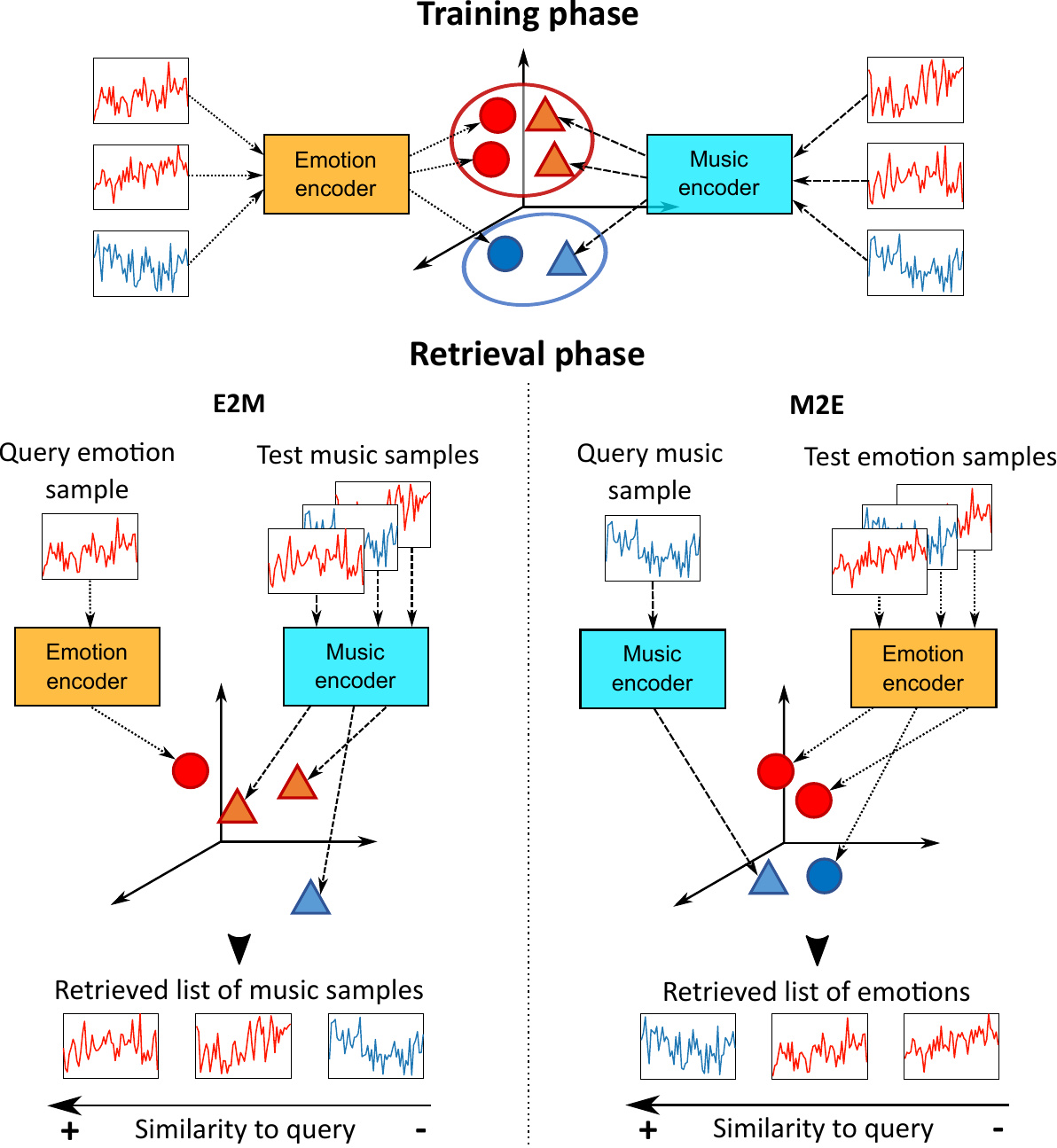}
	\caption{Principle of an embedding-based retrieval approach for MER. A music encoder and an emotion encoder are first jointly trained to project associated music samples and emotions close to each other in an embedding space. The trained encoders are then used to obtain embeddings of the query and test samples. The test samples are finally ordered by decreasing similarity to the query in the embedding space.}
	\label{fig:oneway-CMER}
\end{figure}

In this paper, we propose an \textit{Embedding-based Music Emotion Recognition} (EMER) approach that performs bidirectional retrieval based on the continuous model of emotions. Our approach can directly analyse the similarity between music and emotion in the embedding space, where an embedding designates here the vector representation of a music sample or an emotion in the embedding space. Fig.~\ref{fig:oneway-CMER} shows a standard EMER approach that projects music samples and emotions into an embedding space, in such a way that associated music and emotions are close to each other in the embedding space. This allows it to identify general emotions because similar emotions are gathered close to each other in terms of their embeddings. It can be noted that by projecting highly associated music samples and specific emotions  in proximity, their fine-grained relations are preserved in the embedding space. This way, EMER can treat both of general and specific emotions. Once both music and emotion encoders are trained, they can be used to obtain the embeddings of a query and test samples, and rank the latter by decreasing similarities to the query in the embedding space.

One challenge when working directly with emotional ratings is that they are inherently uncertain, because they are subjectively annotated according to human perceptions which are highly influenced by many factors such as age, personality, cultural background and surrounding conditions~\cite{han2022_survey,Yang_fazzy_class}. We refer to this phenomenon as \textit{emotional uncertainty}. On the one hand, some existing MER approaches bypass this problem by extracting emotion-related information from sources that are more reliable than individual subjective reports, such as music tags~\cite{won2021-tag, ferraro2021} or lyrics~\cite{shang2021, won2021-text}. On the other hand, many other past work directly use subjective emotion ratings without considering emotional uncertainty, and assume that the provided annotations are completely correct. This uncertainty could however cause the trained models to be either inaccurate or biased, and thus negatively impact the recognition performances.

To mitigate the impact of emotional uncertainty, we develop an approach, called \textit{EMER using Composite Loss} (EMER-CL) that trains music and emotion embeddings with a compound loss examining two statistical characteristics that are affected in a limited way by possible inaccuracies in the emotional annotations. Firstly, we assume that even if emotional intensities differ from user to user in terms of intensity values, they remain nevertheless correlated when listening to the same music sample. Thus, Canonical Correlation Analysis (CCA) is used to devise a correlation-based loss~\cite{Deepcca}. This loss enables us to deal with inter-subject variations in emotional intensities by maximising the correlation between music samples and their associated emotions in an embedding space, so as to find their `relative' connection. That is, the embedding space characterises how acoustic features change according to an increase/decrease in arousal/valence intensities and vice versa. Secondly, music samples yielding very different acoustic features can evoke similar emotions. For instance, happiness can be expressed in different genres like rock, blues and jazz. This kind of large intra-class variation in acoustic features for one emotional category makes projecting a music sample or an emotion into a single point (as shown in Fig.~\ref{fig:oneway-CMER}) suboptimal. Thus, we additionally project each of music samples and emotions as a probability distribution in another embedding space~\cite{hama_kl-divergence,Z_Ren}. This idea is implemented by defining a distribution-based loss that measures the Kullback-Leibler (KL) divergence between the probability distribution for a music sample and the one for an emotion in the embedding space.

Our composite loss consisting of the correlation- and distribution-based losses is necessary for managing the aforementioned inter-subject and intra-class variations resulting from the emotional uncertainty. Only using the correlation-based loss cannot cover the large intra-class variation of acoustic features, while the inter-subject variation of emotional intensities cannot be handled only with the distribution-based loss. The experimental results in Section~\ref{subsec:eval_composite} validate the necessity of combining the correlation- and distribution-based losses.

To sum up, this paper contains the following three main contributions: Firstly, we propose EMER-CL that can work with both general and specific emotions since it uses the continuous model of emotions to obtain embeddings of emotions, and the embedding space maintains not only associated music samples and emotions close to each other but also non-associated ones far away. The embedding space serves as a bridge between music samples and emotions and offers bidirectional MER as a by-product of EMER-CL. Secondly, we propose a new composite loss combining the CCA and KL-divergence losses to take into account the emotional uncertainty. Finally, we perform extensive experiments on two benchmark datasets, MediaEval Database for Emotional Analysis in Music~(DEAM)~\cite{DEAM} and PMEmo~\cite{PMEmo}. We demonstrate the effectiveness of EMER-CL over regression baselines not relying on embeddings, of the composite loss over other alternatives and of the features learned by EMER-CL relatively to the state-of-the-art MER methods. In addition, detailed analysis of EMER-CL results reveals that reasonable recognition is robustly attained even in cases of mis-recognition. 

This paper is organised as follows: Section~\ref{sec:related_work} reviews the literature of exsting MER approaches grouped into several categories. Section~\ref{sec:method} details our EMER-CL and Section~\ref{sec:result} reports the experimental results demonstrating its effectiveness. Detailed analysis of recognition results by EMER-CL is conducted in Section~\ref{sec:detail_analysis}. Finally, Section~\ref{sec:conclusion} presents the conclusion and our future work. In addition to these main contents, several appendices are provided to show experimental details, such as hyper-parameter tuning for EMER-CL and the comparison approaches involved in the comparative studies in Section~\ref{sec:result}, additional insights and results for the detailed analysis in Section $5$, and the computational cost of EMER-CL.

\section{Related work}
\label{sec:related_work}

This section provides a short review of existing MER approaches by dividing them into M2E and E2M. We first review existing M2E approaches by classifying them into three categories: ``feature engineering'' that hand-crafts emotion-related acoustic features, ``feature learning'' based on deep learning that automatically learns emotion-related features, and ``relation modelling'' to extract the relationship between emotional intensities and acoustic features obtained by feature engineering or learning. We then discuss past work dealing with E2M. Through this review, we clarify the novelties of the proposed EMER-CL.

\noindent \textbf{Feature engineering for M2E:} Several libraries like MIRtoolbox~\cite{Lartillot_MIRtoolbox_tool} and  openSMILE~\cite{Eyben_Opensmile_tool} are currently available to extract fundamental acoustic features such as Zero-Crossing Rate (ZCR), Root-Mean-Square (RMS) energy, Mel-Frequency Cepstral Coefficients (MFCCs), Short-Time Fourier Transform (STFT), etc. However, acoustic signal analysis alone might not be enough to account for all required acoustic characteristics~\cite{CircularModel_Dufour}. As a result, a large focus of M2E approaches has been put on feature engineering in recent years. Panda {\it et al.}~\cite{MER_AudioFeatures2020_survey} distinguished several types of emotion-related acoustic features including spectral features (low-level feature), rhythm clarity (perceptual feature) and genre (high-level semantic feature). Mo and Niu~\cite{Mo_feat_class} presented an acoustic feature extraction technique that combines three signal processing algorithms, the orthogonal matching pursuit, Gabor functions, and the Wigner distribution function, to provide an adaptive time-varying description of music signals with a higher spatial and temporal resolution. Panda {\it et al.}~\cite{Panda_feat_class} proposed algorithms to extract acoustic features related to musical texture and expressive performance techniques (e.g., vibrato, tremolo and glissando).

The aforementioned work mainly focuses on designing acoustic features and feature selection to effectively estimate an emotion from a music sample, but it can be claimed that feature engineering does not inherently take into account the emotional uncertainty unlike our EMER-CL approach.

\noindent \textbf{Feature Learning for M2E:} The advantage of feature learning is the ability to capture high-level features from raw data or hand-crafted (low-level) features. Feature learning for M2E has become a fast moving research topic due to the increasing interest in deep learning over the past decade. For this reason, we report only the most recent work (i.e. less than five-year-old) that we found related to this topic. Malik {\it et al.}~\cite{Malik_DL_reg} demonstrated the effectiveness of stacking a Convolutional Neural Network (CNN) and Recurrent Neural Network (RNN) to predict arousal/valence from acoustic features exclusively based on log mel-band energy. Dong {\it et al.}~\cite{Dong_DL_reg} developed a Bidirectional Convolutional Recurrent Sparse Network (BCRSN) that uses the spectrogram of audio signals and reduces computational complexity by converting the continuous arousal/valence prediction process to multiple binary classification problems. Sarkar {\it et al.}~\cite{Sarkar_DL_class} applied a CNN taking log-mel spectrogram as input to the four-class classification problem defined by Russell’s model quadrants~\cite{CircularModel_Russell}. Hizlisoy {\it et al.}~\cite{Hizlisoy_DL_class} proposed a Convolutional Long short term memory Deep Neural Network (CLDNN) for the classification of three quadrants excluding low arousal - high valence from Russell’s model quadrants~\cite{CircularModel_Russell}. Choi {\it et al.}~\cite{Choi_DL_transfer} presented a transfer learning approach where a CNN taking mel-spectrograms as input is firstly trained for a music tagging task, and then transferred and fine-tuned for six other tasks such as music genre classification, speech/music classification, emotion prediction etc. Koh {\it et al.}~\cite{koh2021} presented a comparison of state-of-the-art deep feature learning architectures including VGGish and $L^3$-Net that take audio spectrograms as inputs. The two models outperformed MFCCs features on various M2E datasets for either classification or regression tasks. Orjesek {\it et al.}~\cite{orjesek2022} proposed two deep-learning-based architectures to learn features for M2E as a regression problem. The first one is based on a CNN stacked with a bidirectional Gated Recurrent Unit (GRU) and Multi-Layer Perceptron (MLP). The other has the same architecture, except that an autoencoder is inserted between the CNN and bidrectional GRU. The ensemble is trained by adding a reconstruction term to the loss function. He {\it et al.}~\cite{he2022} proposed a two-stage approach for the classification of low/high arousal and valence. Log-mel spectrograms obtained from the raw audio signals are first used to trained a convolutional autoencoder. The encoder is then used to train two bidirectional Long Short-term Memory (LSTM), one for arousal and the other for valence classification.

The aforementioned approaches do not take into account the emotional uncertainty. On the other hand, our EMER-CL approach uses high-level acoustic features learned by a pre-trained VGGish model~\cite{VGGish}, and takes advantage of the composite loss to deal with the emotional uncertainty.

\noindent \textbf{Relation modelling for M2E:} Past work has also investigated relationships between music samples and emotions, although the lower popularity of this topic in  MER research compared to feature engineering or feature learning means that this field is moving at a slower pace. Yang {\it et al.}~\cite{Yang_modeling_reg} built a group-wise MER scheme (GWMER) which divides users into various groups based on user information such as generation, gender, occupation and personality, and trains a Support Vector Regression (SVR) for the prediction of arousal/valence for each group. GWMER can this way partially address the problem that continuous emotions are more affected by subjective issue than discrete emotions when annotating. Yang and Chen~\cite{Yang_modeling_Ranking-based_reg} presented a ranking-based neural network model that ranks a collection of music samples by emotion and determines the emotional intensity of each music sample. Yang and Chen~\cite{Yang_modeling_predDist_reg} and Chin {\it et al.}~\cite{Chin_modeling_reg} developed probabilistic approaches to deal with the emotional uncertainty by estimating the distribution of emotional intensities from hand-crafted acoustic features. Markov and Matsui~\cite{Markov_modeling_reg} showed that modelling with Gaussian Processes (GP) was more powerful than SVR for arousal/valence regression with hand-crafted acoustic features. Fukayama and Goto~\cite{Fukayama_modeling_reg} evaluated the effectiveness of aggregating multiple GP regressions, each trained with different acoustic features. Wang {\it et al.}~\cite{Wang_modeling_reg} presented Acoustic Emotion Gaussians (AEGs) that treat the emotional uncertainty by modelling hand-crafted acoustic features as a parametric probability distribution (soft assignment) instead of a single point (hard assignment). Wang {\it et al.}~\cite{Wang_modeling_histogram_reg} proposed a Histogram Density Mixture (HDM) model that quantises the arousal/valence space into cells and extracts latent histograms representing characteristic emotion distributions over cells based on hand-crafted acoustic features. Wang {\it et al.}~\cite{Wang_modeling_class} developed a MER system for 34 emotional categories based on Hierarchical Dirichlet Process Mixture Model (HDPMM) that links emotion classes using the property of sharing components in the HDPMM.

To the best of our knowledge, relation modelling approaches have so far exclusively relied on feature engineering, and not yet been used with high-level features obtained by feature learning. On the other hand, our EMER-CL approach uses high-level acoustic features based on VGGish, and projects emotional intensities into an embedding space, which enables us to extract high-level feature representations for emotions.

\noindent \textbf{E2M: } E2M has not been explored as extensively as M2E, especially not in the recent MER literature. One possible reason for this scarcity is the fact that the existing acoustic features associated with emotions are high dimensional, and thus not easy to predict directly from emotions using traditional machine learning methods like regression. Except for studies older than a decade~\cite{Kuo_E2M_graph,Ruxanda_E2M_reg,Deng_E2M_reg_recommend,Yang_E2M_MrEmo_reg}, we could find only a single recent method incorporating E2M elements proposed by Deng {\it et al.}~\cite{Deng_E2M_reg_recommend}. In this work, a music recommendation method taking into account the emotions of the user is proposed. An M2E model is first trained using a classification framework to predict the emotional state associated with a music sample. The model is then used to predict an emotional state sequence containing the emotions associated with the last songs listened by the user. A model based on Conditional Random Fields is used to predict the user's current emotion based on this emotional state sequence. Finally, the similarities between the predicted current emotion and the ones associated to songs of the dataset are computed to retrieve relevant songs to be suggested.

In this paper, we propose a new bidirectional MER approach able to perform either M2E and E2M based on projecting music samples and associated emotions in proximity in two embeddings spaces. Unlike Deng {\it et al.}, our approach can directly perform E2M without relying on a M2E system.

\noindent \textbf{Embedding-based recognition: } Approaches in this category have attracted much attention as techniques that can perform effective bidirectional recognition between different modalities (e.g., image, text and audio). When it comes to approaches involving audio modality, the most common investigations include embedding-based recognition between audio and image~\cite{Zeng_closs_S-DCCA_audio-image, Zeng_closs_TNN-C-CCA_audio-image}, between audio and text (lyrics)~\cite{Yu_closs_DCCA_audio-text, won2021-text}, or even between all three of image, audio and text \cite{shang2021}. Some attempts have also focused on extracting meaningful embeddings from music meta-data (e.g. genre, instrument, mood/theme) and playlist information \cite{ferraro2021}. However to the best of our knowledge, no existing work addresses embedding-based recognition between audio and emotion except our previous study~\cite{CMR_takashima}, where MLPs trained with the CCA loss are used to compute embeddings of music samples and emotions. This paper is an extension of our previous study by adopting RNNs in addition to MLPs, devising a composite loss that combines the CCA and KL-divergence losses, and conducing significantly deeper analysis of experimental results.

\section{EMER-CL approach}
\label{sec:method}

\begin{figure}[tbp]
	\centering
	\includegraphics[width=0.8\linewidth]{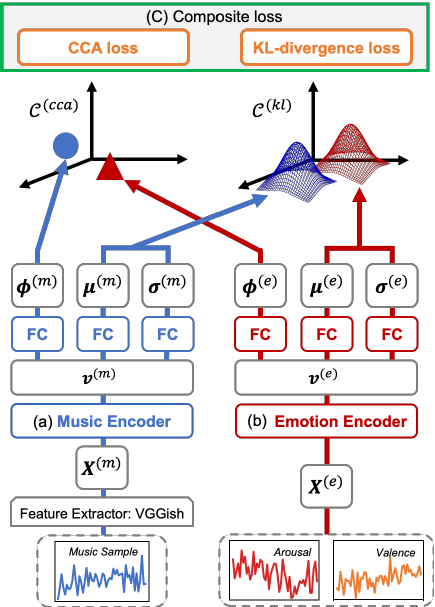} 
	\caption{An overview of our EMER-CL approach. $\boldsymbol{X}^{(m)}$ and $\boldsymbol{X}^{(e)}$ respectively designate the music and emotion sequences input to the music and emotion encoders. $\boldsymbol{v}^{(m)}$ and $\boldsymbol{v}^{(e)}$ are the vectorial outputs of the music and emotion encoders, respectively. Branches of fully connected (FC) layers project $\boldsymbol{v}^{(*)}$ into a point-based embedding $\boldsymbol{\phi}^{(*)}$ in the space $\mathcal{C}^{(cca)}$, and a probabilistic-based embedding following a multivariate Gaussian distribution $\mathcal{N}(\boldsymbol{\mu}^{(*)}, \boldsymbol{\sigma}^{(*)})$ in the space $\mathcal{C}^{(kl)}$, for $* \in \{m,e\}$. Both music and emotion models are jointly trained to minimise a composite loss computing the CCA loss between $\boldsymbol{\phi}^{(m)}$ and $\boldsymbol{\phi}^{(e)}$, and a KL-divergence loss between $\mathcal{N}(\boldsymbol{\mu}^{(m)}, \boldsymbol{\sigma}^{(m)})$ and $\mathcal{N}(\boldsymbol{\mu}^{(e)}, \boldsymbol{\sigma}^{(e)})$.}
	\label{fig:CMER-CL}
\end{figure}

Fig.~\ref{fig:CMER-CL} shows an overview of our EMER-CL approach. First, a music sample is converted into a sequence of acoustic features $\boldsymbol{X}^{(m)} = \boldsymbol{x}_1^{(m)}, \cdots , \boldsymbol{x}_{T_m}^{(m)}$ of length $T_m$. Here, $\boldsymbol{x}_t^{(m)} \in \mathbb{R}^{D_{\boldsymbol{x}}^{(m)}}$ is a $D_{\boldsymbol{x}}^{(m)}$-dimensional feature vector at time $t$ ($1 \leq t \leq T_m$). In our implementation, VGGish~\footnotetext[1]{\url{https://github.com/tensorflow/models/tree/master/research/audioset/vggish}} produced by Google~\cite{VGGish} is used to segment every music sample recorded with a sampling rate of $44.1$kHz into segments of $0.96$ seconds, and then $\boldsymbol{x}_t^{(m)}$ is extracted as a $128$-dimensional vector (i.e., $D_{\boldsymbol{x}}^{(m)} = 128$) from each segment. The emotion associated with the music sample is represented as a sequence $\boldsymbol{X}^{(e)} = \boldsymbol{x}_1^{(e)}, \cdots , \boldsymbol{x}_{T_e}^{(e)}$ of length $T_e$ where $\boldsymbol{x}_t^{(e)} \in \mathbb{R}^{D_{\boldsymbol{x}}^{(e)}}$ is a $D_{\boldsymbol{x}}^{(e)}$-dimensional vector containing characteristics of the emotion at time $t$ ($1 \leq t \leq T_e$). In our setting, $\boldsymbol{x}_t^{(e)}$ is defined as a two-dimensional vector (i.e., $D_{\boldsymbol{x}}^{(e)} = 2$) indicating arousal and valence intensities recorded with a sampling rate of $2$Hz. Unlike $\boldsymbol{X}^{(m)}$, we do not perform feature extraction on the raw arousal/valence intensities and use the latter directly as $\boldsymbol{X}^{(e)}$. This is because we consider that only two types of intensities recorded with a low sampling rate have relatively simple characteristics. Note that for simplicity $\boldsymbol{X}^{(e)}$ is called an \textit{arousal/valence sequence} in the following discussions. 

It is cumbersome to directly project $\boldsymbol{X}^{(m)}$ and $\boldsymbol{X}^{(e)}$ into an embedding space because they are sequences of different lengths $T_m$ and $T_e$. Thus, as shown in Fig.~\ref{fig:CMER-CL} (a) and (b), music and emotion encoders are used to transform $\boldsymbol{X}^{(m)}$ and $\boldsymbol{X}^{(e)}$ into vectors $\boldsymbol{v}^{(m)} \in \mathbb{R}^{D_{\boldsymbol{v}}^{(m)}}$ and $\boldsymbol{v}^{(e)} \in \mathbb{R}^{D_{\boldsymbol{v}}^{(e)}}$, respectively. Each of $\boldsymbol{v}^{(m)}$ and $\boldsymbol{v}^{(e)}$ is a high-level feature that effectively summarises the features in $\boldsymbol{X}^{(m)}$ or $\boldsymbol{X}^{(e)}$ and their temporal relations. We use either an MLP or RNN based on bidirectional Gated Recurrent Unit (GRU)~\cite{Chung_GRU, Schuster_Bidirectional-RNN} as music and emotion encoders as described in Section~\ref{subsec:encoding}.

Then, $\boldsymbol{v}^{(m)}$ and $\boldsymbol{v}^{(e)}$ are projected into an embedding space $\mathcal{C}^{(cca)}$ of dimensionality $D^{(cca)}$ using different Fully Connected (FC) layers with linear activation. The embeddings for $\boldsymbol{v}^{(m)}$ and $\boldsymbol{v}^{(e)}$ in $\mathcal{C}^{(cca)}$ are denoted by $\boldsymbol{\phi}^{(m)} \in \mathbb{R}^{D^{(cca)}}$ and $\boldsymbol{\phi}^{(e)} \in \mathbb{R}^{D^{(cca)}}$, respectively. In addition, two branches of FC layers are used to transform $\boldsymbol{v}^{(m)}$ into a mean vector $\boldsymbol{\mu}^{(m)} \in \mathbb{R}^{D^{(kl)}}$ and a covariance matrix $\boldsymbol{\Sigma}^{(m)} \in \mathbb{R}^{D^{(kl)} \times D^{(kl)}}$. This defines an additional embedding for $\boldsymbol{v}^{(m)}$ as a multivariate Gaussian distribution $\mathcal{N}(\boldsymbol{\mu}^{(m)}, \boldsymbol{\Sigma}^{(m)})$ in another embedding space $\mathcal{C}^{(kl)}$ of dimensionality $D^{(kl)}$. Considering the expensive computational cost to process multivariate Gaussian distributions, we assume that each dimension is independent based on the standard practice of the literature~\cite{Kingma_VAE, Sanchez_GMM}. Thus, $\mathcal{N}(\boldsymbol{\mu}^{(m)}, \boldsymbol{\Sigma}^{(m)})$ is reduced to $\mathcal{N}(\boldsymbol{\mu}^{(m)}, \boldsymbol{\sigma}^{(m)})$ by replacing $\boldsymbol{\Sigma}^{(m)}$ with the variance vector $\boldsymbol{\sigma}^{(m)} \in \mathbb{R}^{D^{(kl)}}$ representing the variance in each dimension. Similarly, $\boldsymbol{v}^{(e)}$ is converted into $\mathcal{N}(\boldsymbol{\mu}^{(e)}, \boldsymbol{\sigma}^{(e)})$ in $\mathcal{C}^{(kl)}$ using two branches of FC layers. Under the above-mentioned setting, EMER-CL trains the music and emotion encoders and the six branches of FC layers by jointly minimising the CCA loss between $\boldsymbol{\phi}^{(m)}$ and $\boldsymbol{\phi}^{(e)}$ and the KL-divergence loss between $\mathcal{N}(\boldsymbol{\mu}^{(m)}, \boldsymbol{\sigma}^{(m)})$ and $\mathcal{N}(\boldsymbol{\mu}^{(e)}, \boldsymbol{\sigma}^{(e)})$, as illustrated in Fig.~\ref{fig:CMER-CL} (c).

The following sections describe encoding of music samples and emotions, and more details of the training process. The dimensionalities $D_{\boldsymbol{v}}^{(m)}$, $D_{\boldsymbol{v}}^{(e)}$, $D^{(cca)}$ and $D^{(kl)}$ are hyper-parameters of EMER-CL whose specific values are provided in Section~\ref{subsec:impl_details}.

\subsection{Music and emotion encoders}
\label{subsec:encoding}

For the music encoder, two types of neural networks were tested: an MLP and an RNN using bidirectional GRU~\cite{Chung_GRU, Schuster_Bidirectional-RNN}. When the former is used, a mean feature vector $\bar{\boldsymbol{x}}^{(m)}$ is computed by averaging $\boldsymbol{x}_1^{(m)}, \cdots , \boldsymbol{x}_{T_m}^{(m)}$ in $\boldsymbol{X}^{(m)}$, and fed into the MLP which performs several non-linear transformations on $\bar{\boldsymbol{x}}^{(m)}$ to output $\boldsymbol{v}^{(m)}$. The RNN using bidirectional GRU computes two types of $D_{\boldsymbol{h}}^{(m)}$-dimensional hidden states $\overrightarrow{\boldsymbol{h}}_t^{(m)} \in \mathbb{R}^{D_{\boldsymbol{h}}^{(m)}}$ and $\overleftarrow{\boldsymbol{h}}_t^{(m)} \in \mathbb{R}^{D_{\boldsymbol{h}}^{(m)}}$ that represent temporal characteristics of $\boldsymbol{X}^{(m)}$ in the forward and backward directions, respectively. Roughly speaking, $\overrightarrow{\boldsymbol{h}}_t^{(m)}$ is computed by recursively aggregating $\boldsymbol{x}_t^{(m)}$ at the current time $t$ and $\overrightarrow{\boldsymbol{h}}_{t-1}^{(m)}$ obtained at the previous time $t-1$. Letting $f$ be a function for the recursive aggregation, $\overrightarrow{\boldsymbol{h}}_t^{(m)}$ is described as $\overrightarrow{\boldsymbol{h}}_t^{(m)} = f (\boldsymbol{x}_t^{(m)}, \overrightarrow{\boldsymbol{h}}_{t-1}^{(m)})$~\cite{Chung_GRU}. In contrast, $\overleftarrow{\boldsymbol{h}}_t^{(m)}$ in the backward direction is computed by aggregating $\boldsymbol{x}_t^{(m)}$ and $\overleftarrow{\boldsymbol{h}}_{t+1}^{(m)}$ at the next time $t+1$, that is, $\overleftarrow{\boldsymbol{h}}_t^{(m)} = f (\boldsymbol{x}_t^{(m)}, \overleftarrow{\boldsymbol{h}}_{t+1}^{(m)})$. The vector obtained by concatenating $\overrightarrow{\boldsymbol{h}}_{T_m}^{(m)}$ and $\overleftarrow{\boldsymbol{h}}_1^{(m)}$ is a high-level feature expressing bidirectional temporal characteristics in $\boldsymbol{X}^{(m)}$, and is fed into the FC layers to produce a higher-level feature as $\boldsymbol{v}^{(m)}$.

Our preliminary experiments showed the effectiveness of an RNN using bidirectional GRU as the emotion encoder regardless of datasets. We hypothesise that this is due the fact that bidirectional GRUs can capture the best the variations in arousal and valence levels during the listening of a music sample, and therefore take this information into account to produce meaningful embeddings. Therefore, a feature vector $\boldsymbol{v}^{(e)}$ for an arousal/valence sequence $\boldsymbol{X}^{(e)}$ is extracted the same way as $\boldsymbol{v}^{(m)}$ when an RNN is used as the music encoder. Finally, the specific values of hyper-parameters like $D_{\boldsymbol{h}}^{(m)}$, $D_{\boldsymbol{h}}^{(e)}$ for the RNN-based emotion encoder, and the configuration of FC layers are provided in Section~\ref{subsec:impl_details}.

\subsection{Training with the composite loss}
\label{subsec:training_encoders}

Let $\mathcal{X} = \{ (\boldsymbol{X}^{(m)}_n, \boldsymbol{X}^{(e)}_n) \}^N_{n=1}$ be a batch consisting of $N$ pairs of an acoustic feature sequence and an arousal/valence sequence for associated music samples and emotions. And, $\mathcal{V} = \{ (\boldsymbol{v}^{(m)}_n, \boldsymbol{v}^{(e)}_n ) \}^N_{n=1}$ is a set of feature pairs obtained by feeding $(\boldsymbol{X}^{(m)}_n, \boldsymbol{X}^{(e)}_n) \in \mathcal{X}$ to the music and emotion encoders. The CCA loss $CCA (\mathcal{F})$ \cite{Deepcca} is computed by converting $\mathcal{V}$ into a set of embeddings $\mathcal{F} = \{(\boldsymbol{\phi}^{(m)}_n,  \boldsymbol{\phi}^{(e)}_n)\}^N_{n=1}$. In addition, the KL-divergence loss $KL( \mathcal{G})$ \cite{hama_kl-divergence} is calculated by transforming $\mathcal{V}$ into a set of pairs of multivariate Gaussian distributions $ \mathcal{G} = \{ ( \mathcal{N}(\boldsymbol{\mu}^{(m)}_n, \boldsymbol{\sigma}^{(m)}_n), \ \mathcal{N}(\boldsymbol{\mu}^{(e)}_n, \boldsymbol{\sigma}^{(e)}_n))\}^N_{n=1}$. Our composite loss $CL(\mathcal{F}, \mathcal{G})$ combines  $CCA(\mathcal{F})$ and $KL( \mathcal{G})$ as follows:
\begin{equation}
	CL(\mathcal{F}, \mathcal{G}) = \lambda \cdot CCA(\mathcal{F}) + (1-\lambda) \cdot KL( \mathcal{G})
	\label{eq:Compositing-CCA-RankLoss} 
\end{equation}
where $\lambda \in [0,1]$ is a weight parameter to balance $CCA(\mathcal{F})$ and $KL( \mathcal{G})$. Details of  $CCA(\mathcal{F})$ and $KL( \mathcal{G})$ are described in the following Sections~\ref{subsubsec:CCALoss} and \ref{subsubsec:RankLoss}.

\subsubsection{Correlation-based embedding with CCA}
\label{subsubsec:CCALoss}

The CCA loss is employed to construct an embedding space $\mathcal{C}^{(cca)}$ where $\{ \boldsymbol{\phi}^{(m)}_n\}_{n=1}^{N}$ and $\{ \boldsymbol{\phi}^{(e)}_n\}_{n=1}^{N}$ are strongly correlated. More specifically, for each dimension of $\mathcal{C}^{(cca)}$, music and emotion embeddings are linearly correlated with each other regardless of their actual values. This linear correlation indicates what change in acoustic features (or emotions) would be associated to the corresponding  change in emotions (or acoustic features) for each of these dimensions. This allows us to characterise the relationship between music samples and emotions in a way that is independent from the actual emotion intensities attributed by individuals, which is useful for addressing the inter-subject variations described in Section~\ref{sec:intro}. To extract embeddings capturing complex correlations between music samples and emotions, FC layers are firstly used to refine $\boldsymbol{v}_n^{(m)}$ and $\boldsymbol{v}_n^{(e)}$ into a $D_{\boldsymbol{z}}^{(m)}$-dimensional vector $\boldsymbol{z}_n^{(m)} \in \mathbb{R}^{D_{\boldsymbol{z}}^{(m)}}$ and a $D_{\boldsymbol{z}}^{(e)}$-dimensional vector $\boldsymbol{z}_n^{(e)} \in \mathbb{R}^{D_{\boldsymbol{z}}^{(e)}}$, respectively. The CCA loss is computed on $\boldsymbol{z}_n^{(m)}$ and $\boldsymbol{z}_n^{(e)}$.

Let $\boldsymbol{z}^{(m)}$ be a random vector that is sampled from the probability distribution estimated using a set of $N$ samples $\{ \boldsymbol{z}_n^{(m)} \}_{n=1}^{N}$, and $\boldsymbol{z}^{(e)}$ be a random vector from the probability distribution estimated using $\{ \boldsymbol{z}_n^{(e)} \}_{n=1}^{N}$. In addition, let us assume that $\boldsymbol{w}^{(m)} \in \mathbb{R}^{D_{\boldsymbol{z}}^{(m)}}$ and $\boldsymbol{w}^{(e)} \in \mathbb{R}^{D_{\boldsymbol{z}}^{(e)}}$ are $D_{\boldsymbol{z}}^{(m)}$- and $D_{\boldsymbol{z}}^{(e)}$-dimensional weight vectors to project $\boldsymbol{z}^{(m)}$ and $\boldsymbol{z}^{(e)}$ into scalars, respectively. CCA optimises $\boldsymbol{w}^{(m)}$ and $\boldsymbol{w}^{(e)}$ so as to maximise the following correlation between $\boldsymbol{w}^{(m)T} \boldsymbol{z}^{(m)}$ and $\boldsymbol{w}^{(e)T} \boldsymbol{z}^{(e)}$~\cite{Deepcca}:
\begin{multline}
	\rho( \boldsymbol{w}^{(m)T} \boldsymbol{z}^{(m)}, \ \boldsymbol{w}^{(e)T} \boldsymbol{z}^{(e)}) \\
	= \frac{
		\boldsymbol{w}^{(m)T} \boldsymbol{\Sigma}^{(me)} \boldsymbol{w}^{(e)}
	}{
		\sqrt{\boldsymbol{w}^{(m)T} \boldsymbol{\Sigma}^{(mm)} \boldsymbol{w}^{(m)}}
		\sqrt{\boldsymbol{w}^{(e)T} \boldsymbol{\Sigma}^{(ee)} \boldsymbol{w}^{(e)}}
	}
	\label{eq: cca-vector}
\end{multline}
where $\boldsymbol{\Sigma}^{(me)} \in \mathbb{R}^{D_{\boldsymbol{z}}^{(m)} \times D_{\boldsymbol{z}}^{(e)}}$ is the cross-covariance matrix computed from $\{ (\boldsymbol{z}^{(m)}_n,\ \boldsymbol{z}^{(e)}_n)\}^N_{n=1}$ and $\boldsymbol{\Sigma}^{(mm)} \in \mathbb{R}^{D_{\boldsymbol{z}}^{(m)} \times D_{\boldsymbol{z}}^{(m)}}$ and $\boldsymbol{\Sigma}^{(ee)} \in \mathbb{R}^{D_{\boldsymbol{z}}^{(e)} \times D_{\boldsymbol{z}}^{(e)}}$ are the covariance matrices for $\{\boldsymbol{z}^{(m)}_n\}^N_{n=1}$ and  $\{\boldsymbol{z}^{(e)}_n\}^N_{n=1}$, respectively. In Eq.~(\ref{eq: cca-vector}) the quantity to maximise is invariant in scaling of $\boldsymbol{w}^{(m)}$ and $\boldsymbol{w}^{(e)}$, so it is possible to focus on the problem where the denominator is equal to 1. In other words, the objective of CCA is to maximise the numerator in Eq.~(\ref{eq: cca-vector}) subject to the constraints  $\boldsymbol{w}^{(m)T}\boldsymbol{\Sigma}^{(mm)}\boldsymbol{w}^{(m)} = 1$ and $\boldsymbol{w}^{(e)T}\boldsymbol{\Sigma}^{(ee)}\boldsymbol{w}^{(e)} = 1$.

The CCA approach described above can be re-applied independently on each dimension of $C^{(cca)}$. For this, $D^{(cca)}$ pairs of weight vectors $\{ ( \boldsymbol{w}^{(m)}_d, \boldsymbol{w}^{(e)}_d)\}_{d=1}^{D^{(cca)}}$ are found to maximise the sum of correlations between $\boldsymbol{w}^{(m)T}_d \boldsymbol{z}^{(m)}$ and $\boldsymbol{w}^{(e)T}_d \boldsymbol{z}^{(e)}$. In other words, letting $\boldsymbol{W}^{(m)} \in \mathbb{R}^{D_{\boldsymbol{z}}^{(m)} \times D^{(cca)}}$ be a matrix where each column is $\boldsymbol{w}^{(m)}_d$, $\boldsymbol{W}^{(m)T} \boldsymbol{z}^{(m)}$ forms a $D^{(cca)}$-dimensional embedding $\boldsymbol{\phi}^{(m)}$. Similarly, $\boldsymbol{W}^{(e)} \in  \mathbb{R}^{D_{\boldsymbol{z}}^{(e)} \times D^{(cca)}}$ where each column is $\boldsymbol{w}^{(e)}_d$ is defined to create $\boldsymbol{\phi}^{(e)} = \boldsymbol{W}^{(e)T} \boldsymbol{z}^{(e)}$. From this perspective, the general CCA maximises the sum of correlations each computed for one dimension of $\boldsymbol{\phi}^{(m)}$ and $\boldsymbol{\phi}^{(e)}$. Past work has shown that the batch optimisation of $\{ ( \boldsymbol{w}^{(m)}_d, \boldsymbol{w}^{(m)}_d)\}_{d=1}^{D^{(cca)}}$ can be done by solving the following constrained optimisation problem~\cite{Deepcca}:
\begin{equation}
	\begin{aligned}
		\text{ minimise : } & CCA(\mathcal{F}) =  
		- \operatorname{tr}(\boldsymbol{W}^{(m)T}\boldsymbol{\Sigma}^{(me)}\boldsymbol{W}^{(e)}) \\
		\text{ subject to : }&\scriptsize{ \boldsymbol{W}^{(m)T}\boldsymbol{\Sigma}^{(mm)}\boldsymbol{W}^{(m)} =\ \boldsymbol{W}^{(e)T}\boldsymbol{\Sigma}^{(ee)}\boldsymbol{W}^{(e)} = \boldsymbol{I}}
	\end{aligned}
	\label{eq:CCA_trace}
\end{equation}
where the trace operation (tr) is used to sum up the correlations on each dimension of $\boldsymbol{\phi}^{(m)}$ and $\boldsymbol{\phi}^{(e)}$. After obtaining the optimal $\boldsymbol{W}^{(m)*}$ and $\boldsymbol{W}^{(e)*}$, the correlation-based embedding for $\boldsymbol{z}^{(m)}_n$ of a music sample and the one for $\boldsymbol{z}^{(e)}_n$ of an emotion are computed as $\boldsymbol{\phi}^{(m)}_n = \boldsymbol{W}^{(m)*T} \boldsymbol{z}^{(m)}_n$ and $\boldsymbol{\phi}^{(e)}_n = \boldsymbol{W}^{(e)*T} \boldsymbol{z}^{(e)}_n$, respectively.

\subsubsection{Distribution-based embedding with KL-divergence}
\label{subsubsec:RankLoss}

The CCA loss analyses only the relation (correlation) between music samples and their associated emotions, but neither the relation of a music sample to non-associated emotions nor the relation of an emotion to non-associated music samples. In addition, $\boldsymbol{\phi}^{(m)}_n$ and $\boldsymbol{\phi}^{(e)}_n$ are points in $\mathcal{C}^{(cca)}$, which is unsuitable for managing the large intra-class variation of acoustic features, as discussed in Section~\ref{sec:intro}. To address these issues, the KL-divergence loss $KL( \mathcal{G})$ is used to build an embedding space $\mathcal{C}^{(kl)}$ that attempts to fulfil the following conditions: 1) a music sample and its associated emotion are projected as multivariate Gaussian distributions $\mathcal{N}(\boldsymbol{\mu}^{(m)}_n, \boldsymbol{\sigma}^{(m)}_n)$ and  $\mathcal{N}(\boldsymbol{\mu}^{(e)}_n, \boldsymbol{\sigma}^{(e)}_n)$ which are similar to each other; 2) a music sample and its non-associated emotion are projected as dissimilar distributions $\mathcal{N}(\boldsymbol{\mu}^{(m)}_n, \boldsymbol{\sigma}^{(m)}_n)$ and  $\mathcal{N}(\boldsymbol{\mu}^{(e)}_{n'}, \boldsymbol{\sigma}^{(e)}_{n'})$ ($n \neq n'$). Similarly, an emotion and its non-associated music sample are transformed into dissimilar distributions $\mathcal{N}(\boldsymbol{\mu}^{(e)}_{n}, \boldsymbol{\sigma}^{(e)}_{n})$ and  $\mathcal{N}(\boldsymbol{\mu}^{(m)}_{n'}, \boldsymbol{\sigma}^{(m)}_{n'})$.

For simplicity, $\mathcal{N}(\boldsymbol{\mu}^{(m)}_{n}, \boldsymbol{\sigma}^{(m)}_{n})$ and $\mathcal{N}(\boldsymbol{\mu}^{(e)}_{n}, \boldsymbol{\Sigma}^{(e)}_{n})$ are abbreviated into $\mathcal{N}_n^{(m)}$ and $\mathcal{N}_n^{(e)}$, respectively. In addition, we use the term \textit{positive pair} to indicate a pair of $\mathcal{N}_n^{(m)}$ and $\mathcal{N}_n^{(e)}$ obtained for a music sample and its associated emotion. On the other hand, a \textit{negative pair} expresses a pair of $\mathcal{N}_n^{(m)}$ and $\mathcal{N}_{n'}^{(e)}$ or a pair of $\mathcal{N}_{n'}^{(m)}$ and $\mathcal{N}_n^{(e)}$, consisting of non-associated music sample and emotion. Note that by following the standard of embedding-based retrieval~\cite{Kiros_triplet-loss,hama_kl-divergence}, we consider that a music sample and an emotion whose indices are the same form a positive pair, and any other pair is a negative pair.

The aforementioned conditions for $\mathcal{C}^{(kl)}$ can be formulated using a triplet $(\mathcal{N}_n^{(m)}, \mathcal{N}_n^{(e)}, \mathcal{N}_{n'}^{(e)})$:
\begin{align}
	\psi \big( \mathcal{N}_n^{(m)}, \mathcal{N}_n^{(e)} \big) \leq \alpha + \psi \big( \mathcal{N}_n^{(m)}, \mathcal{N}_{n'}^{(e)} \big),
	\label{eq:rankloss_inequality}
\end{align}
where $\psi (\cdot,\cdot)$ represents a distance between two distributions. Additionally, $\alpha > 0$ is a margin hyper-parameter which determines how far the difference between the distance for the positive pair $(\mathcal{N}_n^{(m)}, \mathcal{N}_n^{(e)})$ and the one for the negative pair $(\mathcal{N}_n^{(m)}, \mathcal{N}_{n'}^{(e)})$ is allowed to be. Eq.~(\ref{eq:rankloss_inequality}) uses $\mathcal{N}_n^{(m)}$ as an anchor and checks whether its distance to $\mathcal{N}_n^{(e)}$ is sufficiently smaller than its distance to $\mathcal{N}_{n'}^{(e)}$. Similarly, another triplet $(\mathcal{N}_n^{(e)}, \mathcal{N}_n^{(m)}, \mathcal{N}_{n'}^{(m)})$ can define the distance condition using $\mathcal{N}_n^{(e)}$ as an anchor:
\begin{align}
	\psi \big( \mathcal{N}_n^{(e)}, \mathcal{N}_n^{(m)} \big) \leq \alpha + \psi \big( \mathcal{N}_n^{(e)}, \mathcal{N}_{n'}^{(m)} \big).
	\label{eq:rankloss_inequality2}
\end{align}

For each positive pair $(\mathcal{N}_n^{(m)}, \mathcal{N}_{n}^{(e)})$ in $ \mathcal{G}$, $KL( \mathcal{G})$ examines the distance conditions in Eqs.~\ref{eq:rankloss_inequality} and \ref{eq:rankloss_inequality2}. Specifically, the following ranking loss $r (\mathcal{N}_n^{(m)}, \mathcal{N}_n^{(e)})$ is defined by combining two hinge losses as follows:
\begin{align}
	& r ( \mathcal{N}^{(m)}_n, \ \mathcal{N}^{(e)}_n) \label{eq:rankloss_rank} \\
	& = \sum_{n' \neq n} \max \Big\{0, \ \psi \big( \mathcal{N}_n^{(m)}, \mathcal{N}_n^{(e)} \big) - \alpha - \psi \big( \mathcal{N}_n^{(m)}, \mathcal{N}_{n'}^{(e)} \big) \Big\} \nonumber \\
	& + \sum_{n' \neq n}  \max \Big\{0, \ \psi \big( \mathcal{N}_n^{(e)}, \mathcal{N}_n^{(m)} \big) - \alpha - \psi \big( \mathcal{N}_n^{(e)}, \mathcal{N}_{n'}^{(m)} \big) \Big\}, \nonumber
\end{align}
where the first term becomes zero if all the negative pairs defined using $\mathcal{N}^{(m)}_n$ as an anchor lead to distances that are greater than the distance between the positive pair by more than $\alpha$. The second term also checks a similar distance condition using $\mathcal{N}^{(e)}_n$ as an anchor. This way $r ( \mathcal{N}^{(m)}_n, \ \mathcal{N}^{(e)}_n)$ indicates how small $\psi ( \mathcal{N}^{(m)}_n, \ \mathcal{N}^{(e)}_n )$ is relatively to $\psi ( \mathcal{N}^{(m)}_n, \ \mathcal{N}^{(e)}_{n'} )$ and $\psi ( \mathcal{N}^{(m)}_{n'}, \ \mathcal{N}^{(e)}_n )$ defined for all the negative pairs.

To compute $r ( \mathcal{N}^{(m)}_n, \ \mathcal{N}^{(e)}_n)$, the KL-divergence is employed as a distance $\psi (\cdot,\cdot)$ between two multivariate Gaussian distributions $\mathcal{N} ( \boldsymbol{\mu}_1,  \boldsymbol{\sigma}_1 )$ and $\mathcal{N} ( \boldsymbol{\mu}_2,  \boldsymbol{\sigma}_2 )$ in $\mathcal{C}^{(kl)}$ of dimensionality $D^{(kl)}$, and is computed as follows
\small
\begin{align}
	& \psi \big( \mathcal{N}(\boldsymbol{\mu}_1, \boldsymbol{\sigma}_1 ), \ \mathcal{N}(\boldsymbol{\mu}_2, \boldsymbol{\sigma}_2) \big) = \label{eq:KL-divergence-sigma} \\
	& \frac{1}{2} \bigg\{
	\sum_{d=1}^{D^{(kl)}} \frac{\sigma_{1,d}}{\sigma_{2,d}}
	- D^{(kl)}
	+ \ln \frac{ \prod_{d=1}^{D^{(kl)}} \sigma_{2,d}}{ \prod_{d=1}^{D^{(kl)}} \sigma_{1,d}} 
	+ \sum_{d=1}^{D^{(kl)}} \frac{ (\mu_{2,d}-\mu_{1,d})^2 }{ \sigma_{2,d} }
	\bigg\} , \nonumber
\end{align}
\normalsize
where $\boldsymbol{\mu}_1$ and $\boldsymbol{\sigma}_1$ are expanded as $(\mu_{1,1}, \cdots ,  \mu_{1,D^{(kl)}})^T$
and $(\sigma_{1,1}, \cdots ,  \sigma_{1,D^{(kl)}})^T$, respectively. Similarly, $\boldsymbol{\mu}_2$ and $\boldsymbol{\sigma}_2$ are also expanded.

Finally, $KL( \mathcal{G})$ is defined as the sum of $r ( \mathcal{N}^{(m)}_n, \ \mathcal{N}^{(e)}_n)$ for all the positive pairs in $ \mathcal{G}$. The minimisation of $KL( \mathcal{G})$ can therefore lead both music and emotion encoders and FC layers to learn parameters so that the KL-divergence between each positive pair is minimised, while maximising the KL-divergence between each negative pair.

\subsection{Testing EMER-CL in M2E and E2M}
\label{subsec:how_to_recognise}
We evaluate EMER-CL in the framework of M2E and E2M that are formulated as a retrieval task. In the following paragraphs, $q$ designates the index of a query, with $1 \leq q \leq Q$ where $Q$ is the number of examples in the test set. In M2E, the query is $\boldsymbol{X}^{(m)}_q$ which represents a music sample that is associated with an emotion rating $\boldsymbol{X}^{(e)}_q$. The trained music model (consisting in the music encoder and three branches of FC layers) is used to encode the query music sample $\boldsymbol{X}^{(m)}_q$ into its correlation-based embedding $\boldsymbol{\phi}^{(m)}_q$ and distribution-based embedding $\mathcal{N}_q^{(m)}$. Similarly, the trained emotion model (consisting in the emotion encoder and three branches of FC layers) is used to convert the $j$th test emotion $\boldsymbol{X}^{(e)}_j$ (for $1 \leq j \leq Q$) into its correlation-based embedding $\boldsymbol{\phi}^{(e)}_j$ and distribution-based embedding $\mathcal{N}_j^{(e)}$. The similarity $s(\boldsymbol{X}^{(m)}_q,\boldsymbol{X}^{(e)}_j)$ between the query music sample and the $j$th test emotion is computed as follows: 

\small
\begin{align}
	s(\boldsymbol{X}^{(m)}_q,\boldsymbol{X}^{(e)}_j) = \lambda \cdot \Gamma ( \boldsymbol{\phi}_q^{(m)}, \ \boldsymbol{\phi}_j^{(e)} ) + (1-\lambda) \cdot \psi \big( \mathcal{N}_q^{(m)}, \mathcal{N}_j^{(e)} \big),
	\label{eq:similarity_measure}
\end{align}
\normalsize

\noindent where $\lambda \in [0,1]$ is the same weighting parameter used for the loss in Eq.~(\ref{eq:Compositing-CCA-RankLoss}), $\psi$ designates the negative KL-divergence, and $\Gamma$ is a correlation-based similarity between $\boldsymbol{\phi}^{(m)}_q = (\phi_{q,1}^{(m)}, \cdots , \phi_{q,D^{(cca)}}^{(m)})^T$ and $\boldsymbol{\phi}^{(e)}_j = (\phi_{j,1}^{(e)}, \cdots , \phi_{j,D^{(cca)}}^{(e)})^T$. 

To determine a suitable correlation-based similarity $\Gamma$, we proceeded as follows. As a reminder, the CCA loss is designed to maximise the correlation on each of $D^{(cca)}$ dimensions independently. The higher the correlation $\rho_d$ on the $d$-th dimension ($1 \leq d \leq D^{(cca)}$) is, the more linearly aligned embedding values for music samples and emotions in training data (i.e., $\{ \phi_{n,d}^{(m)} = \boldsymbol{w}^{(m)T}_d \boldsymbol{z}_n^{(m)} \}_{n=1}^{N}$ and $\{ \phi_{n,d}^{(e)} = \boldsymbol{w}^{(e)T}_d \boldsymbol{z}_n^{(e)} \}_{n=1}^{N}$) are. Based on this, linear regression is performed to extract the approximation line $g_d(\cdot)$ that takes as input $\phi_{q,d}^{(m)}$ being the embedding value for the query music sample on the $d$th dimension and outputs $g_d(\phi_{q,d}^{(m)})$ being an approximate embedding value for the associated emotion. Thus, the negative of the absolute difference between $g_d(\phi_{q,d}^{(m)})$ and the embedding value $\phi_{j,d}^{(e)}$ for the test emotion is defined as its similarity to the query music sample on the $d$th dimension. By summing up such similarities on all the $D^{(cca)}$ dimensions, $\Gamma ( \boldsymbol{\phi}_q^{(m)}, \boldsymbol{\phi}_j^{(e)} )$ is computed as follows:

\small
\begin{align}
	% Gamma ( \boldsymbol{\phi}_q^{(m)}, \boldsymbol{\phi}_t^{(e)} ) = - \sum_{d = 1 \cap \rho_d \geq P}^{D^{(cca)}} \rho_d  \left| g_d(\phi_{q,d}^{(m)}) - \phi_{t,d}^{(e)} \right| ,
	\Gamma ( \boldsymbol{\phi}_q^{(m)}, & \boldsymbol{\phi}_j^{(e)} ) = \sum_{d \in \{ d' | 1 \leq d' \leq D^{(cca)} \textrm{, } \rho_{d '} \geq P \}} - \rho_d  \left| g_d(\phi_{q,d}^{(m)}) - \phi_{j,d}^{(e)} \right|
	\label{eq:corr_sim}
\end{align}
\normalsize

where each dimension $d'$ is filtered out or weighted by the correlation $\rho_{d'}$. If $\rho_{d'}$ is lower than the threshold $P$, the approximation on the $d'$th dimension is regarded as inaccurate and the similarity on this dimension is not counted. In contrast, as $\rho_{d'}$ becomes higher, the approximation is regarded as more accurate and the similarity is more prioritised by weighting it with $\rho_{d'}$.

The similarities between the query music and all test emotions $s(\boldsymbol{X}^{(m)}_q,\boldsymbol{X}^{(e)}_j)$ are computed for all $1 \leq j \leq Q$ using Eq.~(\ref{eq:similarity_measure}), and then sorted by decreasing similarities. The performance of M2E is evaluated by examining whether the test emotion associated with the query music sample $\boldsymbol{X}^{(e)}_q$ is ranked at a high position in the sorted output list.

Similarly to M2E, E2M is performed by encoding a query emotion $\boldsymbol{X}^{(e)}_q$ and test music samples $\boldsymbol{X}^{(m)}_j$ with the trained emotion and music models, respectively. Then, the $Q$ test music samples are sorted by computing their similarities to the query emotion $s(\boldsymbol{X}^{(e)}_q,\boldsymbol{X}^{(m)}_j)$ for $1 \leq j \leq Q$ according to Eq.~(\ref{eq:similarity_measure}). The rank of the music sample associated with the query emotion is checked to measure the performance of E2M.

\section{Experiments}
\label{sec:result}

In this section, we evaluate EMER-CL on two datasets: MediaEval Database for Emotional Analysis in Music (DEAM)~\cite{DEAM} and PMEmo~\cite{PMEmo}. We first present an overview and pre-processing on each dataset, the evaluation metrics, and the implementation details. Then, we present the results of three experiments. The first is an ablation study to validate the composite loss, the second compares EMER-CL to regression baselines not relying on embeddings, and the last examines the generality of EMER-CL by comparing it to the state-of-the-art MER methods.

\subsection{Datasets}

DEAM~\cite{DEAM}\footnotemark[2] provides $1802$ music samples that are free audio source records, and their corresponding arousal/valence sequences where arousal and valence intensities lie in $[-10, 10]$. Each music sample was annotated with arousal and valence intensities every $0.5$ seconds by at least 5 subjects recruited using Amazon's crowdsourcing platform \textit{Mechanical Turk}. These intensities were projected into the range $[-1, 1]$ for each subject. DEAM contains $1744$ $45$-second-long music samples and $58$ samples that have durations longer than $45$ seconds. The authors of DEAM decided to discard the first $15$ seconds of annotations after observing high instabilities due to a high variance in how music samples start. Because of this and the fact that most music samples last only $45$ seconds, each music sample is normalised to have a length of $30$ seconds by taking the segment starting at $15$ seconds and ending at $45$ seconds. In order to make our system robust for the average music listener, an ``average sequence'' is created for each of arousal and valence by computing the average value over all subjects at each timestamp. The average sequences for arousal and valence are then concatenated into an arousal/valence sequence $\boldsymbol{X}^{(e)}$. Finally, the $30$-second segment corresponding to the paired music sample is extracted. 

\footnotetext[2]{\url{https://cvml.unige.ch/databases/DEAM}}

PMEmo~\cite{PMEmo} (and more specifically the updated dataset PMEmo2019\footnotemark[3]) contains $794$ music samples which are the chorus parts of high quality popular pop-songs gathered from the Billboard Hot $100$, the iTunes Top $100$ Songs (USA) and the UK Top $40$ Singles Chart. 457 subjects including 366 Chinese university students, 44 Chinese music students and 47 English speaking individuals were recruited for the annotation process. Each music sample is annotated with arousal and valence intensities between $1$ (low) and $9$ (high) every $0.5$ seconds, and then projected into the range $[0,1]$. Similarly to DEAM, the first $15$ seconds of annotations were discarded by taking into account a large variance in beginnings of music samples. Unlike DEAM, music samples and associated arousal and valence sequences in PMEmo have variable lengths ranging from $0.08$ to $73.24$ seconds. We decided to select music samples with a total length of at least $7.0$ seconds to evaluate in total $701$ samples. Similarly to DEAM, an arousal/valence sequence $\boldsymbol{X}^{(e)}$ was associated to each music sample by averaging arousal and valence intensities over all subjects who annotated the sample at each timestamp.

\footnotetext[3]{\url{https://github.com/HuiZhangDB/PMEmo}}

\subsection{Evaluation metrics}
\label{evaluation_metrics}

Each dataset is split into training and test partitions with a proportion of $8:2$. Specifically, DEAM is split into training and test partitions, respectively containing $1441$ and $Q = 361$ pairs of a music sample and an emotion. The training and test partitions of PMEmo include $560$ and $Q = 141$ pairs, respectively. A model trained on a training partition is evaluated on the corresponding test partition in the framework of M2E and E2M. On both datasets, M2E is run $Q$ times using each of the $Q$ music samples $\boldsymbol{X}^{(m)}_j$ as a query. For each query, the $Q$ test emotions $\boldsymbol{X}^{(e)}_j$ are sorted in decreasing order of their similarities to the query $s(\boldsymbol{X}^{(m)}_q,\boldsymbol{X}^{(e)}_j)$, and the performance is evaluated by checking the rank of the emotion associated with the query music sample. Similarly, E2M is executed $Q$ times by adopting each of the $Q$ emotions $\boldsymbol{X}^{(e)}_j$ as a query and examining the rank of its associated music sample $\boldsymbol{X}^{(m)}_j$.

%Each dataset is split into training and test partitions with a proportion of $8:2$. Specifically, DEAM is split into training and test partitions containing $1441$ and $361$ pairs of a music sample and an emotion, respectively. Training and test partitions of PMEmo include $560$ and $141$ pairs, respectively. A model trained on a training partition is evaluated on the corresponding test partition in the framework of M2E and E2M. Assuming $Q$ pairs of a music sample and an emotion in the test partition, M2E is run $Q$ times using each of $Q$ music samples as a query. Then, $Q$ emotions are sorted in decreasing order of their similarities to the query, and the performance is evaluated by checking the rank of the emotion associated with the query music sample. Similarly, E2M is executed $Q$ times by adopting each of $Q$ emotions as a query and examining the rank of its associated music sample.

In this framework, only one sample (i.e, emotion or music sample) is associated with a query (i.e., query music sample or query emotion). We use the \textit{Mean Reciprocal Rank} (MRR) as the main evaluation metric as it is commonly employed in retrieval studies~\cite{Yu_closs_DCCA_audio-text,won2021-text,zhao2022}. The MRR is calculated based on $r_q$ that is the rank of the sample associated with a query $q$ in the sorted list of samples ($1 \leq r_q \leq Q$). The MRR is defined as the average of reciprocals of all $r_q$ over $Q$ queries, that is, $\operatorname{MRR} = \frac{1}{|Q|}\sum_{q=1}^{|Q|}\frac{1}{r_q}$. However, the MRR is biased in the sense that it puts much higher priorities on samples ranked at high positions than those at low positions. This can lead to low MRR values even if the rank of the sample associated to the query is low. For example, $r_q=1$ leads to a reciprocal of $1$ while it is close to $0$ ($0.05$) for $r_q=20$, even though $r_q=20$ might still be a very good retrieval result. In our case, samples associated with queries are rarely ranked at the very top positions because each query in either the DEAM or PMEmo dataset has many `close neighbours' (i.e., samples annotated with similar emotions, or from the same music style) that may easily be ranked above the sample associated with the query. The MRR values might therefore be non-intuitive and not trivially interpretable when checking the performances of our system. Thus, we additionally compute the \textit{Average Rank} (AR) that is the average of all $r_q$ over $Q$ queries, namely $\operatorname{AR} = \frac{1}{|Q|}\sum_{q=1}^{|Q|} r_q$. Using the AR, samples can be equally evaluated regardless of their ranks. Although the median of all $r_q$ is one popular evaluation measure for embedding-based retrieval~\cite{Kiros_triplet-loss}, we use their average to be consistent with the calculation of the MRR. To sum up, our evaluation is based on the MRR and AR that respectively become higher and lower as a better performance is obtained. Finally, all models in each configuration are run $10$ times. In each of them, all the parameters in EMER-CL (i.e., parameters of music and emotion encoders and six branches of FC layers in Fig.~\ref{fig:CMER-CL}) are randomly initialised, and a dataset is randomly split into training and test partitions with a proportion of $8:2$. The mean and standard-deviation of MRRs and ARs obtained in $10$ runs are reported.

\subsection{Implementation details}
\label{subsec:impl_details}

We tested various combinations of an MLP and RNN for music and emotion encoders on DEAM and PMEmo. To simplify the selection of encoders, MLPs (or RNNs) with the same architecture were used regardless of encoder types. We found that for the music encoder, an MLP and RNN performed the best on DEAM and PMEmo, respectively. For the emotion encoder, RNNs are the best on both datasets.

The numbers of layers and units per layer of the MLP and RNNs were chosen by grid search. The MLP consists of five FC layers, each of which performs a non-linear transformation based on units using softplus $\sigma(x)=\log(1+e^x)$ as their activation function. The number of units in each layer is $256$ for the first layer, $512$ for the second and third layers, and $1024$ units for the fourth and fifth layers. That is, $D_{\boldsymbol{v}}^{(m)} = 1024$ when the encoder is an MLP. A dropout layer with a dropout rate of $0.5$ is inserted between two consecutive layers.

Each RNN based on bidirectional GRU has a single layer with a $512$-dimensional hidden state (i.e., $D_{\boldsymbol{h}}^{(m)} = D_{\boldsymbol{h}}^{(e)} = 512$), and finally outputs a $1024$-dimensional vector by concatenating the hidden states obtained in the forward and backward directions. This vector is subsequently passed to an MLP consisting of five FC layers where units use softplus as their activation function. The number of units per FC layer was chosen as $512$ for the first three ones, and $1024$ for the two last ones (i.e., $D_{\boldsymbol{v}}^{(m)} = D_{\boldsymbol{v}}^{(e)} = 1024$). A dropout layer with a dropout rate of $0.5$ is also added behind all layers except the RNN layer and the output layer.

Regarding the embedding spaces based on the CCA and KL-divergence losses, their dimensionalities are set to  $D^{(cca)} = D^{(kl)} = 1024$. Based on our experiments, we recommend to set the threshold for the correlation-based similarity $P$, the margin in the KL-divergence loss $\alpha$ and the combination weight for the composite loss $\lambda$ to default values of $(P,\alpha,\lambda) = (0.4,1.0,0.5)$ when testing our approach on a new dataset. It is nevertheless possible to optimise these values on a specific dataset by following the optimisation strategy detailed in Appendix~\ref{sec:tuning_CMER-CL}. The results in the next subsections were obtained using the optimal hyper-parameters $(P,\alpha,\lambda) = (0.5,1.0,0.6)$ on DEAM and $(P,\alpha,\lambda) = (0.7,2.0,0.5)$ on PMEmo.

Our EMER-CL model is trained using Adam \cite{adam_optimisation} as the optimiser with an initial learning rate of $1e^{-5}$. The model was trained for $5001$ and $10001$ epochs on DEAM and PMEmo, respectively. We implemented all the codes using TensorFlow library (version $1.15$) on a machine equipped with Intel i9-9900K CPU, $64$GB RAM, NVIDIA RTX 2080Ti GPU and CUDA version $10.0$.

\subsection{Evaluation of the composite loss}
\label{subsec:eval_composite}

To evaluate the effectiveness of our proposed composite loss (\textit{Composite}), we compare its performance to the ones individually obtained only using the CCA loss (\textit{CCA-Loss}) or KL-divergence loss (\textit{KL-Loss}). In addition, to examine the effectiveness of projecting music samples and emotions as probability distributions, we implement the most popular embedding approach that projects them as points based on their cosine similarities in an embedding space ~\cite{Kiros_triplet-loss}. For this approach, all the configurations of our EMER-CL model are the same except that $\boldsymbol{v}^{(m)}$ and $\boldsymbol{v}^{(e)}$ from the music and emotion encoders are projected into vectors instead of multivariate Gaussian distributions, and $\psi(\cdot,\cdot)$ in Eq.~(\ref{eq:rankloss_rank}) is replaced with the negative of their cosine similarity. We report both the performance only using the loss based on cosine similarities (\textit{Cos-Loss}) and the one obtained by the composite loss combining \textit{CCA-Loss} and \textit{Cos-Loss} (\textit{Composite-C}). Following the optimisation strategy described in Appendix~\ref{sec:tuning_composite_c}, the hyper-parameters of \textit{Cos-Loss} and \textit{Composite-C} were also optimised to $(P,\alpha,\lambda) = (0.9,0.3,0.7)$ on DEAM and $(P,\alpha,\lambda) = (0.9,0.1,0.1)$ on PMEmo

Table~\ref{tab:superiority_CL} shows the results obtained with the five losses described previously. For both DEAM and PMEmo, the MRRs and ARs using \textit{KL-Loss} are better than those using \textit{CCA-Loss}. This may be due to the fact that \textit{CCA-Loss} alone only considers the correlation between music samples and emotions, and does not necessarily make sure that music samples and emotions in positive pairs are placed close to each other in the embedding space. On the other hand, \textit{KL-Loss} assigns music samples and emotions in positive pairs to similar multivariate Gaussian distributions while distinguishing the ones in negative pairs by dissimilar distributions. Furthermore, the performances are significantly improved when using \textit{Composite} compared to using only \textit{KL-Loss} or \textit{CCA-Loss}. This verifies the effectiveness of \textit{Composite} that simultaneously considers \textit{CCA-Loss} and \textit{KL-Loss}. Also, the fact that \textit{KL-Loss} outperforms \textit{Cos-Loss} and \textit{Composite} outperforms \textit{Composite-C} on both DEAM and PMEmo datasets indicates the effectiveness of distribution-based embeddings compared to point-based ones.

\begin{table}[tbp]
	\scriptsize
	\setlength{\tabcolsep}{3.5pt}
	\renewcommand{\arraystretch}{1.25}
	\centering
	\caption{Comparison of MRRs and ARs using five different losses. The average and standard deviation of each metric over 10 runs are reported on both datasets.}
	\label{tab:superiority_CL}
	\vspace{-5pt}
	\begin{tabular*}{\columnwidth}{@{\extracolsep{\fill}}l cc cc @{\extracolsep{\fill}}}
		\multicolumn{5}{c}{\textbf{DEAM}} \\ \midrule
		\multirow{2}{*}{LossType} & \multicolumn{2}{c}{M2E}   & \multicolumn{2}{c}{E2M} \\  \cmidrule(lr){2-3} \cmidrule(lr){4-5} 
		& MRR & AR & MRR & AR \\ \midrule
		\textit{CCA-Loss}   & $0.048\pm0.007$ & $85.4\pm5.1$ & $0.046\pm0.006$  & $86.9\pm4.3$  \\
		\textit{KL-Loss}  & $0.071\pm0.007$  & $69.9\pm5.2$ & $0.069 \pm 0.006$ & $69.9 \pm 5.2$ \\ 
		\textit{Cos-Loss}  & $0.060\pm0.006$  & $70.7\pm4.4$ & $0.069 \pm 0.006$ & $70.7\pm 4.0$ \\
		\textit{Composite} & $\mathbf{0.128\pm0.011}$ & $\mathbf{56.0\pm1.4}$ & $\mathbf{0.114\pm0.011}$  &          $\mathbf{55.3\pm1.8}$           \\
		\textit{Composite-C} & $0.063\pm0.010$ & $70.5\pm5.4$ & $0.078\pm0.010$  &  $68.4\pm2.5$           \\
		\bottomrule
		\multicolumn{5}{c}{ } \\ 
		\multicolumn{5}{c}{\textbf{PMEmo}} \\ \midrule    
		\multirow{2}{*}{LossType} & \multicolumn{2}{c}{M2E}   & \multicolumn{2}{c}{E2M} \\  \cmidrule(lr){2-3} \cmidrule(lr){4-5} 
		& MRR & AR & MRR & AR \\ \midrule
		\textit{CCA-Loss}   & $0.063\pm0.008$ & $44.3\pm2.3$ & $0.071\pm0.011$  & $43.8\pm2.3$          \\
		\textit{KL-Loss}  & $0.448\pm0.112$ & $5.0\pm2.4$ & $0.442\pm0.113$  & $5.0\pm2.4$          \\
		\textit{Cos-Loss}  & $0.099\pm0.013$  & $37.6\pm2.8$ & $0.093 \pm 0.010$ & $37.2 \pm 2.6$ \\
		\textit{Composite} & $\mathbf{0.544\pm0.056}$ & $\mathbf{3.8\pm0.6}$ & $\mathbf{0.542\pm0.051}$  & $\mathbf{3.8\pm0.5}$          \\
		\textit{Composite-C} & $0.365\pm0.054$ & $9.5\pm1.7$ & $0.361\pm0.056$  &  $9.6\pm1.7$           \\
		\bottomrule
	\end{tabular*}
	\normalsize 
	% \vspace{-0.3cm}
\end{table}

Finally, it can be noted that the performances on PMEmo are significantly better than those on DEAM. This could be attributed to the fact that music samples in PMEmo are more standardised, for instance by including only chorus parts of pop songs, which leads the music encoder to find more specialised feature. In other words, the higher diversity in music samples on DEAM makes M2E and E2M on this dataset likely to be more difficult.

\subsection{Comparison with the baseline models}
\label{subsec:comparison_baseline}

To the best of our knowledge, no existing EMER method that can be directly compared to EMER-CL has been proposed yet. In addition, all the existing approaches using continuous emotion modelling only perform M2E based on a regression approach to predict real-valued characteristics of the arousal/valence sequence (e.g., the average arousal or valence) for a given music sample~\cite{Schmidt_DL_reg, Weninger_DL_reg, Li_DL_DBLSTM_reg, Malik_DL_reg, Dong_DL_reg, Yang_modeling_reg, Yang_modeling_Ranking-based_reg, Yang_modeling_predDist_reg, Markov_modeling_reg, Fukayama_modeling_reg, Wang_modeling_reg, Wang_modeling_histogram_reg}. Moreover, no existing method can handle E2M to predict acoustic features of the music sample for a given arousal/valence sequence. Considering the aforementioned state of the current MER research, we define the following regression-based baselines to show the effectiveness of EMER-CL.

\noindent \textbf{M2E baselines:} Two M2E baselines \textit{RegBiGRU-M2E} and \textit{RegMLP-M2E} train a regression model that analyses a query music sample and outputs a two-dimensional emotion vector $\bar{\boldsymbol{x}}'^{(e)}$ representing the arousal and valence averaged over time for this query music sample. In particular, \textit{RegBiGRU-M2E} predicts $\bar{\boldsymbol{x}}'^{(e)}$ by applying an RNN based on bidirectional GRU to a sequence of acoustic features $\boldsymbol{X}^{(m)}$, while \textit{RegMLP-M2E} employs an MLP that uses the mean $\bar{\boldsymbol{x}}^{(m)}$ of features in $\boldsymbol{X}^{(m)}$ over time to compute $\bar{\boldsymbol{x}}'^{(e)}$. Both baselines are trained to minimise the Mean Absolute Error (MAE) between $\bar{\boldsymbol{x}}'^{(e)}$ and the ground-truth mean emotion $\bar{\boldsymbol{x}}^{(e)}$ computed from the actual arousal/valence sequence $\boldsymbol{X}^{(e)}$.

In the evaluation, given the $q$th test music sample as a query, we evaluate the trained model by predicting its mean emotion $\bar{\boldsymbol{x}}'^{(e)}_q$ and checking whether $\bar{\boldsymbol{x}}'^{(e)}_q$ is similar to the ground-truth mean emotion $\bar{\boldsymbol{x}}^{(e)}_q$. To this end, we compute the similarities of $\bar{\boldsymbol{x}}'^{(e)}_q$ to the ground-truth mean emotions $\{ \bar{\boldsymbol{x}}^{(e)}_t \}_{t=1}^{Q}$ for all the $Q$ test music samples. The Absolute Error (AE) between $\bar{\boldsymbol{x}}'^{(e)}_q$ and $\bar{\boldsymbol{x}}^{(e)}_t$ is used as their dissimilarity. Then, the ground-truth mean emotions are sorted in ascending order of their AEs to get the rank $r_q$ of $\bar{\boldsymbol{x}}^{(e)}_q$. Finally, $r_q$ is used to calculate an MRR and AR.

\noindent \textbf{E2M baselines:} Similarly to the M2E baselines, an RNN based on bidirectional GRU (\textit{RegBiGRU-E2M}) and an MLP model (\textit{RegMLP-E2M}) are used as E2M baselines to predict a mean acoustic feature $\bar{\boldsymbol{x}}'^{(m)}$. \textit{RegBiGRU-E2M} and \textit{RegMLP-E2M} take as input an arousal/valence sequence $\boldsymbol{X}^{(e)}$ and the mean vector $\bar{\boldsymbol{x}}^{(e)}$ of $\boldsymbol{X}^{(e)}$, respectively. \textit{RegBiGRU-E2M} and \textit{RegMLP-E2M} are trained to minimise the MAE between $\bar{\boldsymbol{x}}'^{(m)}$ and the ground-truth mean acoustic feature $\bar{\boldsymbol{x}}^{(m)}$ computed from  a sequence of acoustic features $\boldsymbol{X}^{(m)}$. Like M2E, an MRR and AR is computed by predicting $\bar{\boldsymbol{x}}'^{(m)}_q$ for the $q$th test emotion, measuring the AEs of $\bar{\boldsymbol{x}}'^{(m)}_q$ to the ground-truth mean acoustic features $\{ \bar{\boldsymbol{x}}^{(m)}_t \}_{t=1}^{Q}$ for all the $Q$ music samples, and check the rank $r_q$ of $\bar{\boldsymbol{x}}^{(m)}_q$.

The baseline models were tuned by grid search to find the hyper-parameters leading to the best performances on each dataset. Details about this hyper-parameter tuning can be found in Appendix~\ref{sec:tuning_baselies}. Table~\ref{tab:superiority_CMER-CL} shows the comparison between the above-mentioned baselines and EMER-CL (referred to as \textit{Composite} in Table~\ref{tab:superiority_CL}). EMER-CL significantly outperforms the baselines based on one-way regression of emotion or acoustic features. This highlights the superiority of our embedding-based approach over traditional regression methods not using embeddings.

\begin{table}[tbp]
	\setlength{\tabcolsep}{2.3pt}
	\renewcommand{\arraystretch}{1.25}
	\centering
	\caption{Comparison between the baselines and EMER-CL (\textit{Composite}). The average and standard deviation of each metric over 10 runs are reported on both datasets.}
	\label{tab:superiority_CMER-CL}
	% \vspace{-5pt}
	\begin{tabular}{c cc cc}
		\multicolumn{5}{c}{\textbf{DEAM}} \\
		\midrule
		\multirow{2}{*}{ModelType} & \multicolumn{2}{c}{M2E}   & \multicolumn{2}{c}{E2M} \\  \cmidrule(lr){2-3} \cmidrule(lr){4-5} 
		&\multicolumn{1}{c}{MRR} & \multicolumn{1}{c}{AR} & \multicolumn{1}{c}{MRR} & \multicolumn{1}{c}{AR} \\ \midrule
		\textit{RegMLP} & $0.066\pm0.002$ & $82.5\pm0.3$  & $0.039\pm0.0003$  & $136.5\pm0.1$ \\
		\textit{RegBiGRU} & $0.057\pm0.005$ & $83.2\pm2.7$  & $0.046\pm0.001$ & $131.2\pm0.3$          \\
		\textit{Composite} & $\mathbf{0.128\pm0.011}$ & $\mathbf{56.0\pm1.4}$ & $\mathbf{0.114\pm0.011}$  &          $\mathbf{55.3\pm1.8}$ \\
		\bottomrule
		\multicolumn{5}{c}{ } \\ 
		\multicolumn{5}{c}{\textbf{PMEmo}} \\ \midrule
		\multirow{2}{*}{ModelType} & \multicolumn{2}{c}{M2E}   & \multicolumn{2}{c}{E2M} \\  \cmidrule(lr){2-3} \cmidrule(lr){4-5} 
		&\multicolumn{1}{c}{MRR} & \multicolumn{1}{c}{AR} & \multicolumn{1}{c}{MRR} & \multicolumn{1}{c}{AR} \\ \midrule
		\textit{RegMLP} & $0.082\pm0.010$ & $38.4\pm1.5$  & $0.050\pm0.002$ & $67.3\pm0.5$ \\
		\textit{RegBiGRU} & $0.095\pm0.011$ & $38.2\pm0.8$  & $0.067\pm0.003$ & $57.3\pm0.2$ \\
		\textit{Composite} & $\mathbf{0.544\pm0.056}$ & $\mathbf{3.8\pm0.6}$ & $\mathbf{0.542\pm0.051}$  & $\mathbf{3.8\pm0.5}$          \\
		\bottomrule
	\end{tabular}
	% \vspace{-0.3cm}
\end{table}

\subsection{Comparison to the MER state-of-the-art}

To evaluate the features learnt by EMER-CL, we performed a comparison of the latter against the state-of-the-art on DEAM and PMEmo. To the best of our knowledge, the EMER problem remains still unexplored in the literature, which makes it difficult to find a past study with which our results can be directly compared. Therefore, we carry out the evaluation on the significantly more popular task of MER. 

The MER literature is fairly scattered, with each study carrying out experiments on different datasets, choosing different evaluation metrics and strategies. Our experiments on both DEAM and PMEmo were carried out based on the most commonly used setting, that is, a K-fold cross validation evaluated either using the Root Mean Squared Error (RMSE), the Pearson correlation coefficient $R$ or the coefficient of determination $R^2$.

Our EMER-CL model is trained for $K=10$ folds using the default values of $\alpha=1$ and $\lambda=0.5$ ($P$ is only required for a retrieval problem, not a regression one). After training the whole of our EMER-CL model, three vectors obtained from the music model (i.e., $\boldsymbol{\phi}^{(m)}$, $\boldsymbol{\mu}^{(m)}$ and $\boldsymbol{\sigma}^{(m)}$ in Fig.~\ref{fig:CMER-CL}) are used to train a soft-margin Support Vector Regression model (C-SVR) with Radial Basis Function (RBF) kernel that attempts to predict the arousal and valence intensities associated with a music sample that is input to the music model. Three different configurations for the input to C-SVR are tested: $\boldsymbol{\phi}^{(m)}$ used alone (also referred to as \textit{EMER-CL (cca)}), $\boldsymbol{\mu}^{(m)}$ and $\boldsymbol{\sigma}^{(m)}$ concatenated together (\textit{EMER-CL (kl)}), and all the three vectors concatenated together (\textit{EMER-CL (all)}). On DEAM, the target to predict was chosen as $\bar{\boldsymbol{x}}^{(e)}$, the average of an arousal/valence sequence $\boldsymbol{X}^{(e)}$ over time. On PMEmo, the last arousal and valence values of the emotion sequence $\boldsymbol{X}^{(e)}$ were predicted instead. The hyper-parameters of the C-SVR (soft-margin and kernel parameters) were optimised by maximising the average $R^2$ after grid search.

\begin{table}[tbp]
	\setlength{\tabcolsep}{3.3pt}
	\centering
	\caption{Comparison between EMER-CL and the state-of-the-art for M2E. The metrics for EMER-CL are provided as the average obtained on 10 folds.}
	\label{tab:m2e_results}
	% \vspace{-5pt}
	\begin{tabular}{c c c}
		\multicolumn{3}{c}{\textbf{DEAM}} \\ \midrule
		Approach & $R^2$ Arousal & $R^2$ Valence \\  
		\midrule
		Hult et al. \cite{hult2020} & $0.35$ & $0.34$ \\
		Cheuk et al. \cite{cheuk2020} & $\mathbf{0.672}$ & $0.367$ \\
		%Bathigama et al. \cite{}  & $\mathbf{0.1501}$ &  $~0.45$ & $0.1682$ & $~0.25$\\ 
		\midrule
		\textit{EMER-CL (cca)} & $0.484$ & $0.661$ \\ 
		\textit{EMER-CL (kl)} & $0.491$ & $\mathbf{0.667}$ \\ 
		\textit{EMER-CL (all)} & $0.489$ & $0.664$ \\ 
		\bottomrule
	\end{tabular}
	\begin{tabular}{c ccc ccc}
		\multicolumn{7}{c}{ } \\ 
		\multicolumn{7}{c}{\textbf{PMEmo}} \\ \midrule
		\multirow{2}{*}{Approach} & \multicolumn{3}{c}{Arousal}   & \multicolumn{3}{c}{Valence} \\  \cmidrule(lr){2-4} \cmidrule(lr){5-7} 
		&\multicolumn{1}{c}{RMSE}  & \multicolumn{1}{c}{$R$} & \multicolumn{1}{c}{$R^2$} & \multicolumn{1}{c}{RMSE} & \multicolumn{1}{c}{$R$} & \multicolumn{1}{c}{$R^2$} \\ \midrule
		Zhang et al. \cite{PMEmo} & $0.102$ & $0.764$ & - & $0.124$ & $0.638$ & - \\
		Hult et al. \cite{hult2020} & - & - & $\mathbf{0.64}$ & - & - & $0.42$ \\
		Chapaneri et al. \cite{chapaneri2020} & $\mathbf{0.064}$ & - & $0.618$ & $\mathbf{0.093}$ & - & $0.376$\\
		de Berardinis et al. \cite{deberardinis2020} & $0.232$ & - & $0.600$ & $0.232$ & - & $0.481$\\\midrule
		\textit{EMER-CL (cca)} & $0.115$ & $0.701$ & $0.493$ & $0.111$ & $0.792$ & $0.628$ \\ 
		\textit{EMER-CL (kl)} & $0.115$ & $0.702$ & $0.494$ & $0.111$ & $\mathbf{0.793}$ & $0.629$ \\ 
		\textit{EMER-CL (all)} & $0.115$ & $0.702$ & $0.495$ & $0.111$ & $\mathbf{0.793}$ & $\mathbf{0.630}$ \\ 
		\bottomrule
	\end{tabular}
	% \vspace{-0.3cm}
\end{table}

Table~\ref{tab:m2e_results} shows the results obtained for arousal and valence prediction on DEAM and PMEmo. The learnt features $\boldsymbol{\phi}^{(m)}$, $\boldsymbol{\mu}^{(m)}$ and $\boldsymbol{\sigma}^{(m)}$ can compete with the state-of-the-art for MER. This indicates that our EMER-CL model can still yield proper MER performances. Curiously, $\boldsymbol{\phi}^{(m)}$, $\boldsymbol{\mu}^{(m)}$ and $\boldsymbol{\sigma}^{(m)}$ yield average results for arousal prediction, and notably good ones for valence prediction which is commonly considered as the most difficult of the two problems. All the three tested combinations of $\boldsymbol{\phi}^{(m)}$, $\boldsymbol{\mu}^{(m)}$ and $\boldsymbol{\sigma}^{(m)}$ return fairly similar performances.

\section{Detailed analysis}
\label{sec:detail_analysis}

MRRs and ARs are global metrics that only depend on the rank $r_q$ of the music sample or emotion associated with a query. It is also desirable to check the relevance of the top-ranked music samples (or emotions) to the query. For this, we compute for M2E an \textit{average cosine similarity} that averages the cosine similarities between the mean acoustic feature of the query music sample $\bar{\boldsymbol{x}}^{(m)}_{q}$ and the ones associated with the top $5\%$ emotions retrieved by EMER-CL, i.e., $\{\bar{\boldsymbol{x}}^{(m)}_{j} | 1 \leq j \leq Q \textrm{, } r_j \leq 0.05 \times Q \}$. For E2M, this average cosine similarity is computed in a likewise way by taking the mean of the cosine similarities computed between the query emotion averaged over time $\bar{\boldsymbol{x}}^{(e)}_{q}$ and the ones associated with the top $5\%$ retrieved music samples in E2M, i.e., $\{\bar{\boldsymbol{x}}^{(e)}_{j} | 1 \leq j \leq Q \textrm{, } r_j \leq 0.05 \times Q \}$. A high average cosine similarity means that EMER-CL can recognise music samples that express emotions similar to a query emotion, or emotions expressed in music samples which are acoustically similar to a query music sample.

\begin{figure}[tbp]
	\centering
	\includegraphics[width=\linewidth]{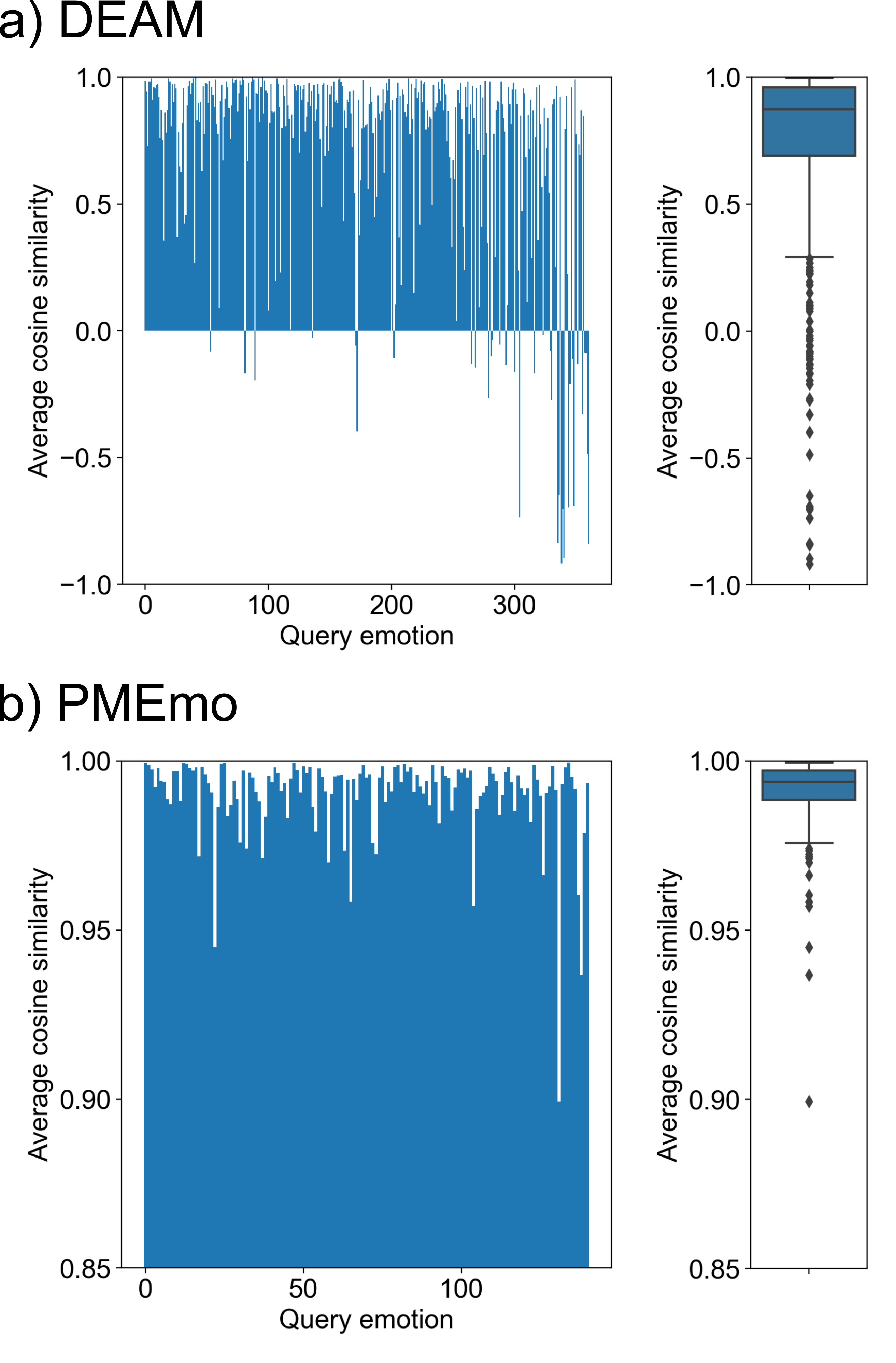}
	\caption{Bar graphs showing average cosine similarities and their box plots on DEAM and PMEmo.}
	\label{fig:detailed_sim_onebar}
	% \vspace{-0.5cm}
\end{figure}

In what follows, we present the analysis for E2M since the mean emotion for each music sample is two-dimensional and can be interpreted easily. It should be noted that arousal/valence intensities are in $[-1, 1]$ and $[0, 1]$ for DEAM and PMEmo respectively, meaning that average cosine similarities range between -1 and 1 on DEAM, and 0 and 1 on PMEmo. Fig.~\ref{fig:detailed_sim_onebar} shows the average cosine similarities for DEAM and PMEmo. In the bar graphs in the left side of Fig.~\ref{fig:detailed_sim_onebar} (a) and (b), each query emotion on the horizontal axis is sorted in increasing order of the rank $r_q$ of its associated music sample. That is, the more to the left a query emotion is, the higher its ground truth music sample was ranked for E2M, meaning that the music sample associated with the query emotion was well recognised.

As shown in the bar graphs of Fig.~\ref{fig:detailed_sim_onebar} (a) and (b), the average cosine similarities obtained on both DEAM and PMEmo are fairly high (close to 1), showing that the top $5\%$ recognised music samples are relevant to most query emotions regardless of the ranks of their ground-truth music samples. More specifically, the median of these average cosine similarities is $0.873$ for DEAM and $0.994$ for PMEmo (the reason for the very high average cosine similarities on PMEmo is provided in Appendix~\ref{sec:detailed_pmemo}). In addition, the box plots in the right side of Fig.~\ref{fig:detailed_sim_onebar} (a) and (b)  show the variations in the average cosine similarities. Here, at least 75 percent of all the average cosine similarities are higher than the 25th percentile (first quartile). The fact that the 25th percentile for DEAM and PMEmo are $0.692$ and $0.989$ respectively, indicates that EMER-CL is able to robustly recognise music samples associated to highly similar emotions to a query emotion. In other words, even if the music sample associated to a query emotion was ranked at a low position, the top 5\% music samples recognised by EMER-CL still exhibit emotions close to the query emotion. Nevertheless, average cosine similarities for some query emotions in DEAM are low, indicating room for improvement in the future.

The same experiment for M2E that computes the average cosine similarity between the acoustic feature of a query music sample and those of music samples associated with the top 5\% emotions in M2E showed similarly good performances. Figures showing such average cosine similarities can be found in Appendix~\ref{sec:detailed_m2e}, and the medians of average cosine similarities on DEAM and PMEmo are $0.753$ and $0.837$, respectively.

%A similar study by computing the average cosine similarity between the acoustic feature of a query music sample and those of music samples associated with the top 5\% emotions in M2E showed similarly good performances. Figures showing such average cosine similarities can be found in Appendix~\ref{sec:detailed_m2e}, and the medians of average cosine similarities on DEAM and PMEmo are $0.753$ and $0.837$, respectively.

\footnotetext[4]{\url{https://mu-lab.info/naoki_takashima/emer-cl}}

\begin{figure*}[tbp]
	\centering
	\includegraphics[width=\linewidth]{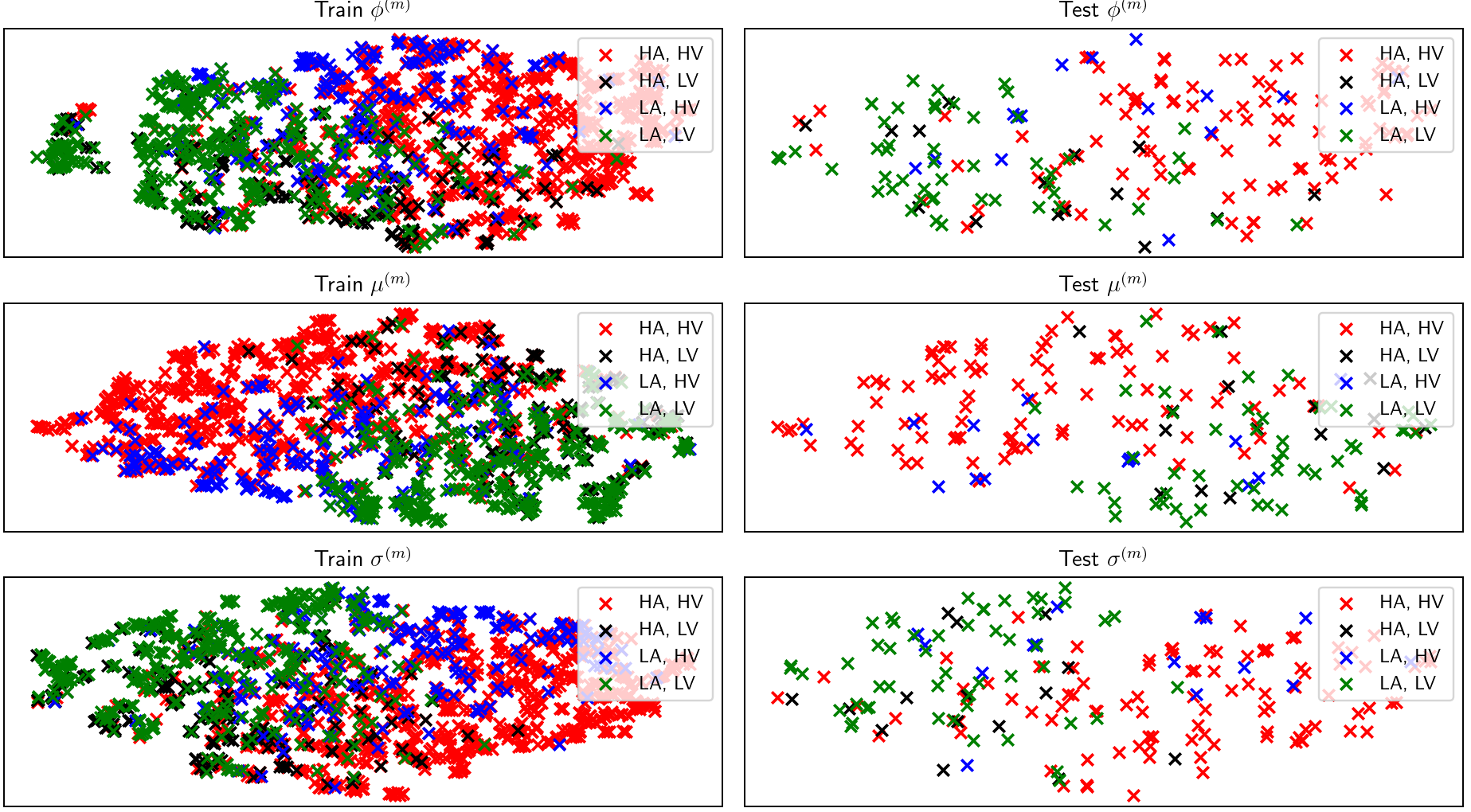}
	\caption{t-SNE projections of the music embeddings $\phi^{(m)}$, $\mu^{(m)}$ and $\sigma^{(m)}$ (respectively first, second and third rows) on the DEAM training (left column) and testing (right column) sets. The t-SNE projections are labelled with their associated emotional quadrants, i.e., high arousal/high valence (HA, HV), high arousal/low valence (HA, LV), low arousal/low valence (LA, LV) and low arousal/high valence (LA, HV).}
	\label{fig:tsne-deam-music}
	% \vspace{-0.5cm}
\end{figure*}

\begin{figure*}[tbp]
	\centering
	\includegraphics[width=\linewidth]{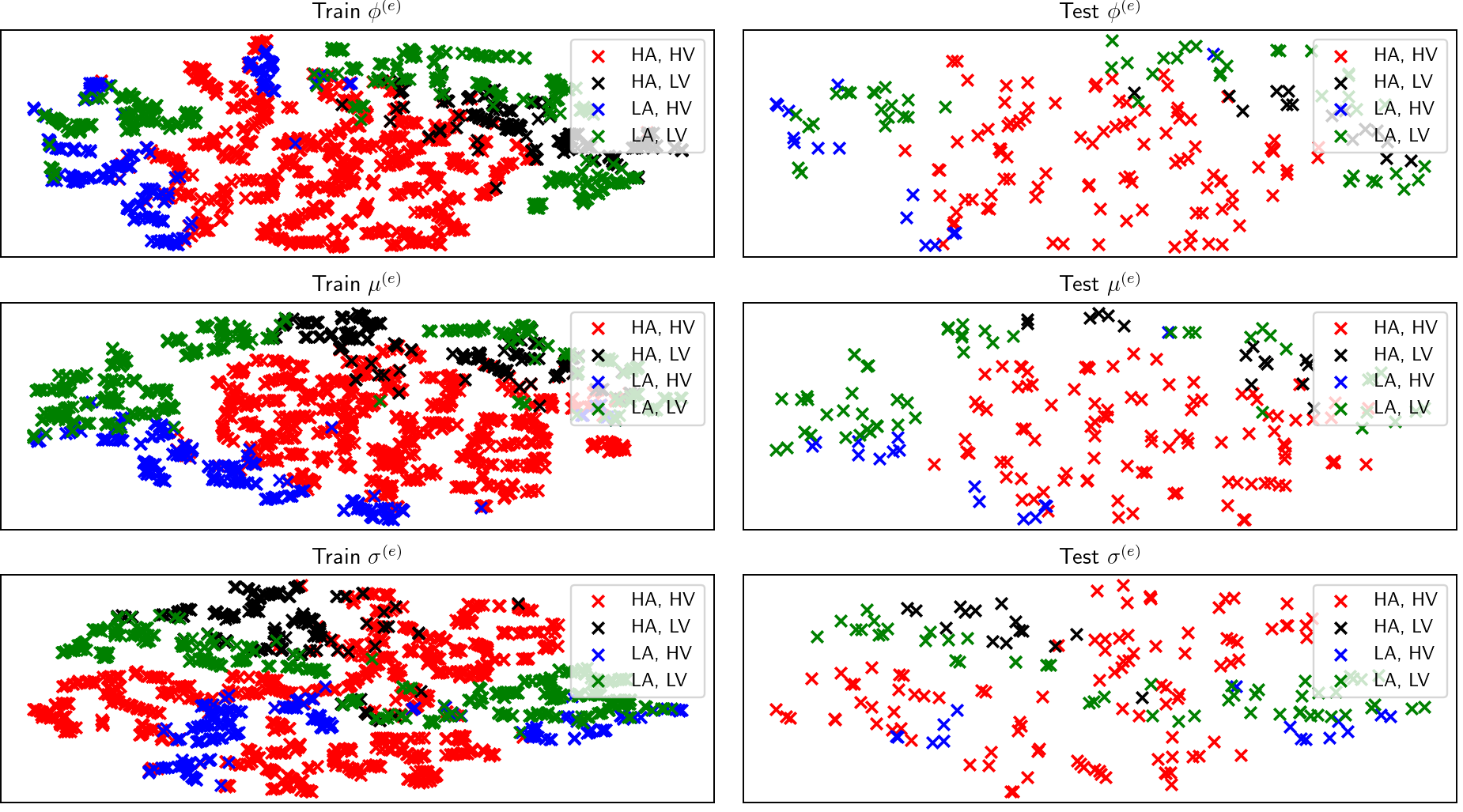}
	\caption{t-SNE projections of the emotion embeddings $\phi^{(e)}$, $\mu^{(e)}$ and $\sigma^{(e)}$ (respectively first, second and third rows) on the DEAM training (left column) and testing (right column) sets. The t-SNE projections are labelled with their associated emotional quadrants, i.e., high arousal/high valence (HA, HV), high arousal/low valence (HA, LV), low arousal/low valence (LA, LV) and low arousal/high valence (LA, HV).}
	\label{fig:tsne-deam-emotion}
	% \vspace{-0.5cm}
\end{figure*}

Finally as a last check of the validity of our approach, we also visualise the embeddings learnt by our music and emotion models on the training and testing sets of both DEAM and PMEmo using t-SNE~\cite{tsne}. We plotted the correlation-based and probabilistic-based embeddings produced by the music model ($\boldsymbol{\phi}^{(m)}$, $\boldsymbol{\mu}^{(m)}$ and $\boldsymbol{\sigma}^{(m)}$) and the emotion model ($\boldsymbol{\phi}^{(e)}$, $\boldsymbol{\mu}^{(e)}$ and $\boldsymbol{\sigma}^{(e)}$). The t-SNE projections were labelled with the emotion-related information available on both DEAM and PMEmo datasets as follows: each emotion sequence $\boldsymbol{X}^{(e)}_j$ in either the training or testing set was first averaged over time to obtain the two-dimensional emotion vector $\bar{\boldsymbol{x}}^{(e)}_{j}$, and then associated to one of the four quadrants of the arousal/valence space: high arousal/high valence (HA/HV), high arousal/low valence (HA/LV), low arousal/low valence (LA/LV) and low arousal/high valence (LA/HV). The t-SNE projections of the embeddings obtained either from $\boldsymbol{X}^{(e)}_j$ or its associated music sample $\boldsymbol{X}^{(m)}_j$ were then annotated with one of these four labels. The t-SNE plots for the DEAM dataset are provided in Figs.~\ref{fig:tsne-deam-music} and~\ref{fig:tsne-deam-emotion} for the music and emotion embeddings, respectively. Since DEAM emotion annotations lie in the range $[-1,1]$, the cut-off value for the definition of the quadrants was set to $0$ for both arousal and valence.

It can be seen from both Figs.~\ref{fig:tsne-deam-music} and~\ref{fig:tsne-deam-emotion} that the DEAM music and emotion embeddings associated with opposite emotional quadrants (e.g. HA/HV and LA/LV, or HA/LV and LA/HV) are very well separated in their respective embedding spaces for both the training and testing examples. This indicates that our EMER-CL approach was successful in learning to project music samples associated to similar emotions close to each other in the embedding space, while simultaneously maximising the distance between embeddings of dissimilar emotions. Similar plots can be obtained for the PMEmo dataset and are provided in Appendix~\ref{sec:tsne-pmemo}.

\section{Conclusion and Future Work}
\label{sec:conclusion}

In this paper, we introduced an Embedding-based Music Emotion Recognition using Composite Loss~(EMER-CL) approach that projects music samples and emotions into embedding spaces, in order to consider both general emotional categories and fine-grained discrimination within each category. In particular, to deal with the emotional uncertainty, EMER-CL uses the composite loss consisting of the CCA loss to maximise the correlation between music samples and their associated emotions in an embedding space, and the KL-divergence loss to project them as similar multivariate Gaussian distributions in another embedding space. The experimental results on DEAM and PMEmo validate the effectiveness of the composite loss, embedding-based approach and features learned by EMER-CL. In addition, a detailed analysis of EMER-CL's results demonstrates that it can robustly recognise reasonable music samples (or emotions) even when failing to identify the ground-truth ones.

To further improve the performance of EMER-CL, we aim to extend the music and emotion encoders by pre-training them with self-supervised learning~\cite{contrastive_self-supervised_learning_survey} which can learn underlying feature representations using unlabelled data. We also plan to adopt a self-attention layer~\cite{attention_is_all_you_need} which can capture long-term dependencies of features, and have led to promising performances when jointly used with bidirectional LSTM and GRU, in particular for sentiment analysis~\cite{basiri2021} or sensor-based emotion recognition~\cite{tao2020}. Because emotions are also strongly dependent on cultural background, we also plan to use MER datasets that provide detailed background information about their raters in future work. The generality of our approach across cultures could be then be demonstrated.

Finally, the codes (and the instruction of data usage) used in this paper are available on our GitLab repository\footnotemark[4], in order for other researchers to reproduce the results and extend the current EMER-CL more easily.

\appendices

\newpage

\section{Hyper-parameter Tuning for EMER-CL}
\label{sec:tuning_CMER-CL}

This section presents how to tune EMER-CL's hyper-parameters, especially $P \in [0,1]$ used in the correlation-based similarity to filter out useless dimensions characterised by weak correlations between music samples and their associated emotions, $\alpha \in \mathbb{R}^{+}$ used in the KL-divergence loss to handle the margin between associated music-emotion pairs and non-associated ones, and $\lambda \in [0,1]$ to control the combination weights of the CCA and KL-divergence losses. Only using either of these losses, an embedding space can be constructed to perform M2E and E2M. Thus, $P$ is firstly tuned based on the performances of M2E and E2M only using the CCA loss. Similarly, $\alpha$ is tuned by carrying out M2E and E2M only with the KL-divergence loss. Finally, $\lambda$ is tuned based on M2E and E2M by combining the CCA and KL-divergence losses that are configured by the separately optimised $P$ and $\alpha$, respectively.

\subsection{Tuning $P$}
\label{sec:tuning_P}

\begin{figure*}[htbp]
	\centering
	\includegraphics[width=\linewidth]{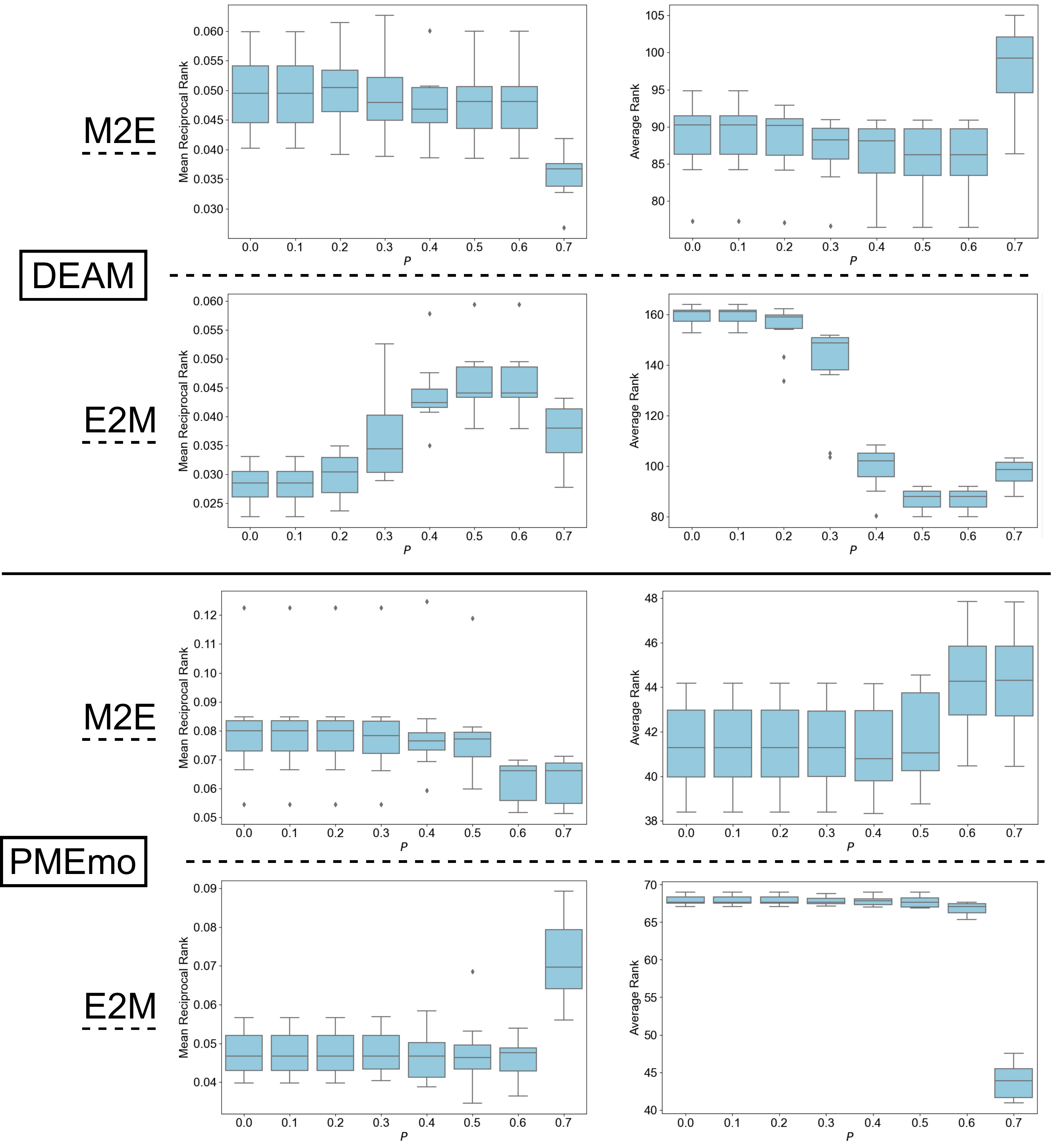}
	\caption{Transitions of EMER-CL performances (MRRs and ARs) obtained only using the CCA loss that is configured by different values of $P$. The optimal $P$ value is selected as $0.5$ and $0.7$ for DEAM and PMEmo, respectively.}
	\label{fig:tuning_P}
\end{figure*}

Fig.~\ref{fig:tuning_P} shows the various performances obtained only using the CCA loss configured by different values of $P$. As previously described in Section~\ref{sec:result}, a performance is measured by an MRR and AR. A good performance is indicated by a high MRR and a low AR. Fig.~\ref{fig:tuning_P} shows the box plots of the MRR and AR performances obtained for each value of $P$ over $10$ runs. They were computed as follows:  Since building an embedding space based on the CCA loss is independent of the choice of $P$, $10$ spaces are firstly constructed by randomly initialising all parameters in EMER-CL (i.e., the parameters of both music and emotion encoders and the ones of six branches of FC layers in Fig.~\ref{fig:CMER-CL}\footnote{Since the KL-divergence loss is always zero in this setting, the four branches of FC layers to produce mean and variance vectors of multivariate Gaussian distributions are not trained.}), and randomly splitting a dataset into training and test partitions with a proportion fixed to $8:2$. Then, every value of $P$ is used to filter out useless dimensions in each of these $10$ embedding spaces to get $10$ performances. 

To determine a range of values to be tested for $P$, we check the maximum correlations among the $1024$ dimensions of each embedding space. In particular, among the $10$ embedding spaces constructed for each dataset, the maximum correlation is $0.783$-$0.833$ and $0.719$-$0.774$ for DEAM and PMEmo, respectively\footnote{When the composite loss is used, the maximum correlations for $10$ CCA-based embedding spaces increase to $0.856$-$0.857$ and $0.9995$-$0.9997$ for DEAM and PMEmo, respectively. This is possibly due to more generalised music and emotion encoders being trained by exploiting both the CCA and KL-divergence losses, which leads to higher-quality embedding spaces. Thus, better EMER-CL's performances than those reported in this paper might be obtained by carrying out grid search on $P$, $\alpha$ and $\lambda$. But, due to its expensive computational cost, we opt to separately tune these hyper-parameters in this paper.}. We test values between $0$ and the maximum correlation with increments of $0.1$ for $P$.

In each graph of Fig.~\ref{fig:tuning_P}, the larger $P$ is, the smaller the number of dimensions used in the correlation-based similarity is. In other words, if $P$ is large, only a small number of dimensions characterised by correlations higher than it are used to compute correlation-based similarities. For each dataset, the optimal value of $P$ is selected as the one that yields the best `overall' performance by considering both M2E and E2M performances. We provide an example of how to select the optimal $P$ value on DEAM by referring to Fig.~\ref{fig:tuning_P}. As seen from this figure, although $P=0.2$ yields the highest median of MRRs in M2E, $P=0.5$ and $0.6$ both lead to the highest median of MRRs in E2M and the lowest medians of ARs in both M2E and E2M. Thus, $P=0.5$ or  $0.6$ can be considered as optimal on DEAM. In this case in particular, $P=0.5$ is selected after observing that its neighbouring value $P=0.4$ yields higher performances than $P=0.7$ neighbouring $P=0.6$. For PMEmo, $P=0.7$ is chosen because of its significantly higher performances on E2M compared to the other $P$ values. It can be noted that the performances of \textit{CCA-Loss} in the comparative study in Section~\ref{subsec:eval_composite} are nothing but the ones that are obtained only using the CCA loss based on the aforementioned optimal $P$ values.

\subsection{Tuning $\alpha$}
\label{sec:tuning_alpha}

\begin{figure*}[htbp]
	\centering
	\includegraphics[width=\linewidth]{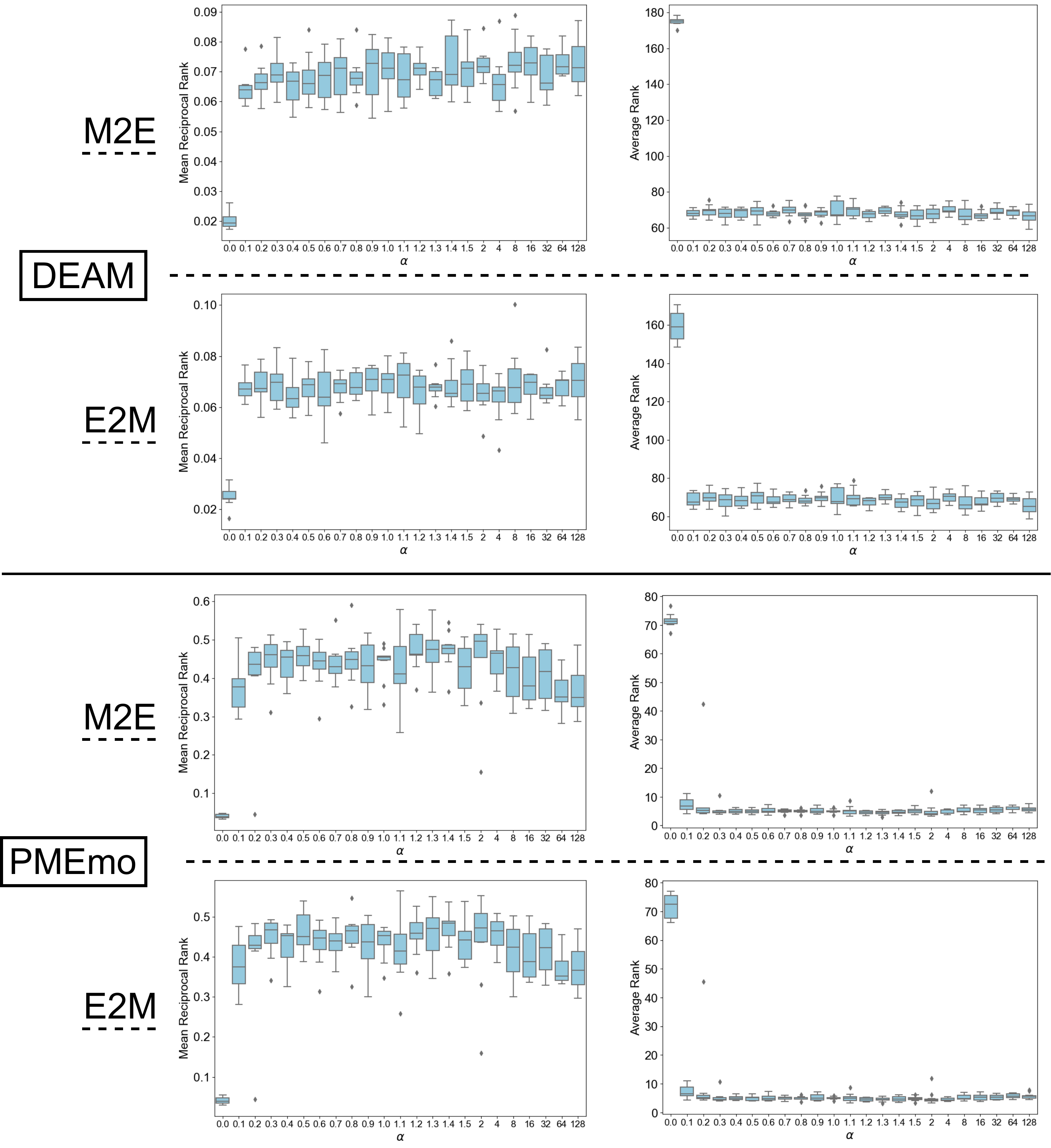}
	\caption{Transitions of EMER-CL performances (MRRs and ARs) obtained only using the KL-divergence loss that is configured by different values of $\alpha$. The optimal $\alpha$ value is selected as $1.0$ and $2.0$ for DEAM and PMEmo, respectively.}
	\label{fig:tuning_alpha}
\end{figure*}

Unlike $P$ that is bounded, the margin $\alpha$ can theoretically take any positive value. We decided to test values between $0$ and $1.5$ with increments of $0.1$, and powers of $2$ between $2$ and $128$. Using the same box plot format as Fig.~\ref{fig:tuning_P}, Fig.~\ref{fig:tuning_alpha} illustrates the performances obtained using only the KL-divergence loss, which is configured by different values of $\alpha$. Using a similar strategy as for $P$, we select the optimal $\alpha$ value in Fig.~\ref{fig:tuning_alpha} as the one that leads to the best overall performance. As \textit{it} can be seen from Fig.~\ref{fig:tuning_alpha}, \textit{the} performances are similar for all tested values, with the exception of small values $0.0$, $0.1$ and $0.2$ for which the performances are significantly worse. We decide to select $\alpha=1.0$ and $2.0$ as the optimal values for DEAM and PMEmo respectively, on the basis that the overall performances with these $P$ values are slightly higher than those with the others. It can be noted that the performances acquired by the aforementioned optimal $\alpha$ values are reported as the ones of \textit{KL-Loss} in the comparative study in Section~\ref{subsec:eval_composite}.

\subsection{Tuning $\lambda$}
\label{sec:tuning_lambda}

\begin{figure*}[htbp]
	\centering
	\includegraphics[width=\linewidth]{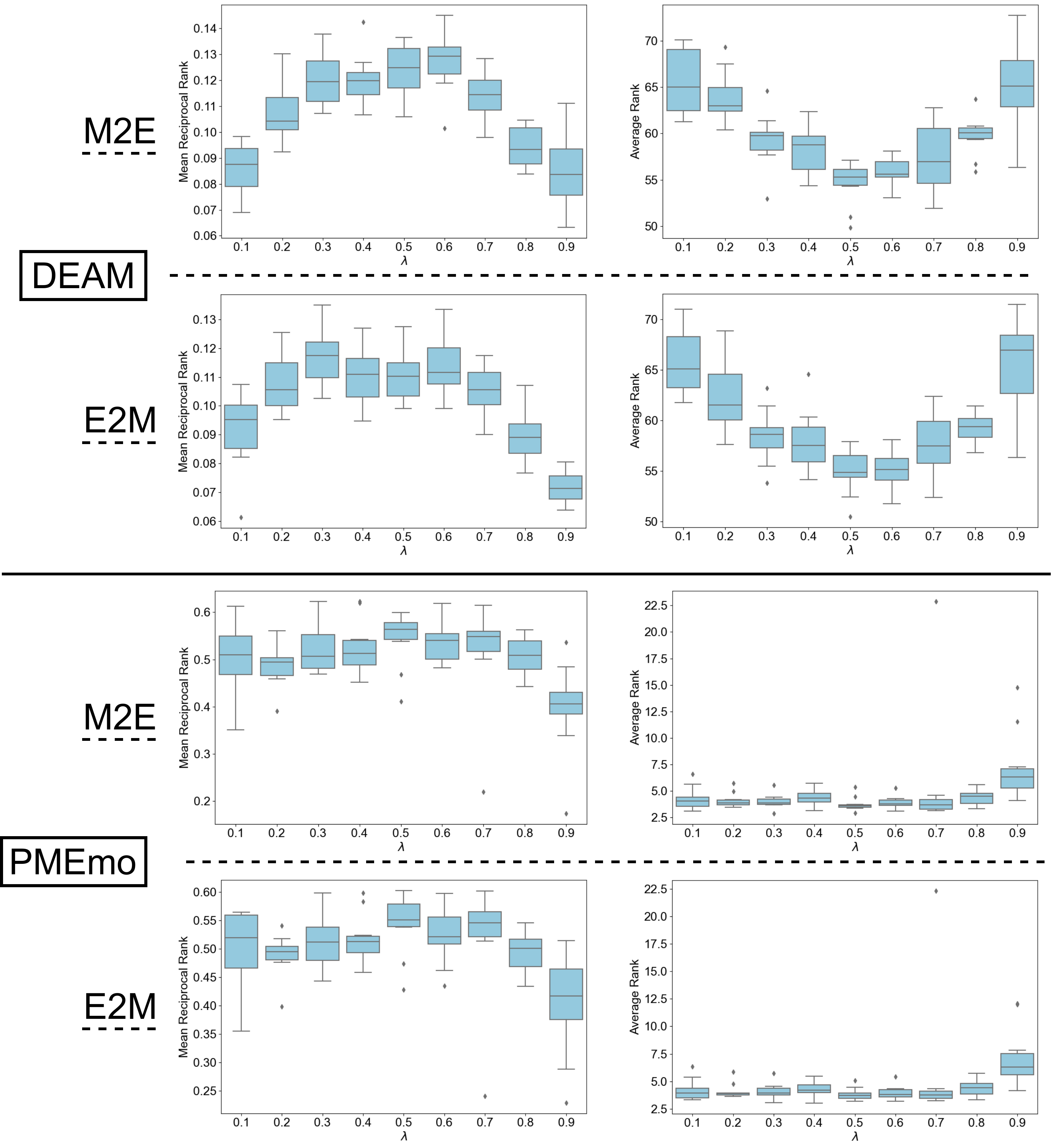}
	\caption{Transitions of EMER-CL performances (MRRs and ARs) obtained by different $\lambda$ values, each of which is used to combine the CCA and KL-divergence losses configured by the optimal $P$ and $\lambda$ values found from Figs.~\ref{fig:tuning_P} and \ref{fig:tuning_alpha}, respectively. The optimal $\lambda$ value is selected as $0.6$ and $0.5$ for DEAM and PMEmo, respectively.}
	\label{fig:tuning_lambda}
\end{figure*}

We tested values of $\lambda$ between $0$ and $1$ with increments of $0.1$. In the same manner as Figs.~\ref{fig:tuning_P} and \ref{fig:tuning_alpha}, Fig.~\ref{fig:tuning_lambda} illustrates EMER-CL's performances obtained for different values of $\lambda$. For each of them, the CCA and KL-divergence losses that are configured by the optimal $P$ and $\alpha$ values (found from Figs.~\ref{fig:tuning_P} and \ref{fig:tuning_alpha} respectively) are used to compute the composite loss. The higher $\lambda$ is, the higher the weight of the CCA loss is. In particular, $\lambda=0$ means only using the KL-divergence loss while only the CCA loss is used with $\lambda=1$. Following the same strategy as the ones employed for choosing the optimal $P$ and $\lambda$ values, $\lambda=0.6$ is chosen as the optimal value yielding the best overall performance on DEAM, and similarly $\lambda=0.5$ is regarded as optimal for PMEmo.

To summarise the whole EMER-CL hyper-parameter selection process, the optimal values found are $P=0.5$, $\alpha=1.0$ and $\lambda=0.6$ on DEAM, and $P=0.7$, $\alpha=2.0$ and $\lambda=0.5$ on PMEmo. The performances of EMER-CL using these optimal values are reported as the ones of \textit{Composite} in Section~\ref{subsec:eval_composite}. In addition, it should be noted that these optimal hyper-parameter values only marginally differ from the default ones (i.e., $P=0.4$, $\alpha=1.0$ and $\lambda=0.5$). Table~\ref{tab:CMER-CL_comp_default} shows the performances of EMER-CL with optimised parameters - referred to as \textit{Composite} - and the ones obtained with the default parameters - referred to as \textit{Composite-d}. As \textit{it} can be seen from this table, the performances of the latter are relatively similar to the ones of the former. The marginal difference in hyper-parameter values between \textit{Composite} and \textit{Composite-d} and their similar performances validate the relevance of the default values.

\begin{table}[!htbp]
	\setlength{\tabcolsep}{2.3pt}
	\renewcommand{\arraystretch}{1.25}
	\centering
	\caption{Performance comparison between EMER-CL using the optimal hyper-parameter values (\textit{Composite}) and EMER-CL using the default ones (\textit{Composite-d})}
	\label{tab:CMER-CL_comp_default}
	\vspace{-5pt}
	\begin{tabular}{c cc cc}
		\multicolumn{5}{c}{\textbf{DEAM}} \\
		\midrule
		\multirow{2}{*}{ModelType} & \multicolumn{2}{c}{M2E}   & \multicolumn{2}{c}{E2M} \\  \cmidrule(lr){2-3} \cmidrule(lr){4-5} 
		&\multicolumn{1}{c}{MRR} & \multicolumn{1}{c}{AR} & \multicolumn{1}{c}{MRR} & \multicolumn{1}{c}{AR} \\ \midrule
		\textit{Composite} & $0.128\pm0.011$ & $56.0\pm1.4$ & $0.114\pm0.011$  &          $55.3\pm1.8$ \\
		\textit{Composite-d} & $0.112\pm0.001$ & $57.4\pm3.2$ & $0.108\pm0.006$  &          $56.5\pm3.6$ \\
		\bottomrule
		\multicolumn{5}{c}{ } \\ 
		\multicolumn{5}{c}{\textbf{PMEmo}} \\ \midrule
		\multirow{2}{*}{ModelType} & \multicolumn{2}{c}{M2E}   & \multicolumn{2}{c}{E2M} \\  \cmidrule(lr){2-3} \cmidrule(lr){4-5} 
		&\multicolumn{1}{c}{MRR} & \multicolumn{1}{c}{AR} & \multicolumn{1}{c}{MRR} & \multicolumn{1}{c}{AR} \\ \midrule
		\textit{Composite} & $0.544\pm0.056$ & $3.8\pm0.6$ & $0.542\pm0.051$  & $3.8\pm0.5$          \\
		\textit{Composite-d} & $0.538\pm0.037$ & $4.0\pm0.5$ & $0.497\pm0.037$  & $4.5\pm0.6$          \\
		\bottomrule
	\end{tabular}
\end{table}

\section{Hyper-parameter Tuning for \textit{Composite-C}}
\label{sec:tuning_composite_c}

In this section, we tune hyper-parameters of the two methods \textit{Cos-loss} and \textit{Composite-C} used in the comparative study in Section~\ref{subsec:eval_composite}. \textit{Cos-Loss} produces point-based embeddings using the loss based on cosine similarities between music samples and emotions~\cite{Kiros_triplet-loss}, and is used to examine the effectiveness of our proposed \textit{KL-Loss} implementing distribution-based embeddings. Both of \textit{Cos-Loss} and \textit{KL-Loss} construct an embedding space in the same ranking loss framework that involves $\alpha$ to control the margin between associated music-emotion pairs and non-associated ones. Thus, $\alpha$ for \textit{Cos-Loss} is tuned in the same manner as $\alpha$ for \textit{KL-Loss} in Section~\ref{sec:tuning_alpha}.

Similarly to \textit{Composite}, \textit{Composite-C} is characterised by $\lambda$ that handles the combination weights of \textit{CCA-Los} and \textit{Cos-Loss}. However, while \textit{Cos-Loss} is based on normalised cosine similarities ranging from $-1$ to $1$, the KL-divergences used in \textit{KL-Loss} are contained in a much wider range, like on average about $5.1$ for DEAM and $423.7$ for PMEmo. It should be noted that $P$ impacts the range of correlation-based similarities because it determines the number of dimension-wise similarities counted to compute an overall similarity, as seen from Eq. (\ref{eq:corr_sim}). Thus, both $P$ and $\lambda$ can be tuned to balance the combination of \textit{CCA-Loss} and \textit{Cos-Loss}, and the $P$ values found in Fig.~\ref{fig:tuning_P} for \textit{Composite} may not be suitable for \textit{Composite-C}. Consequently, after selecting the optimal $\alpha$ for \textit{Cos-Loss}, grid search on $P$ and $\lambda$ is carried out to avoid missing an effective combination of \textit{CCA-Loss} and \textit{Cos-Loss} for \textit{Composite-C}. This grid search for \textit{Composite-C} is more exhaustive and favourable than the separate optimisation of $P$ and $\lambda$ employed for \textit{Composite}.

\subsection{Tuning $\alpha$ of \textit{Cos-Loss}}
\label{sec:tuning_alpha_c}

\begin{figure*}[htbp]
	\centering
	\includegraphics[width=\linewidth]{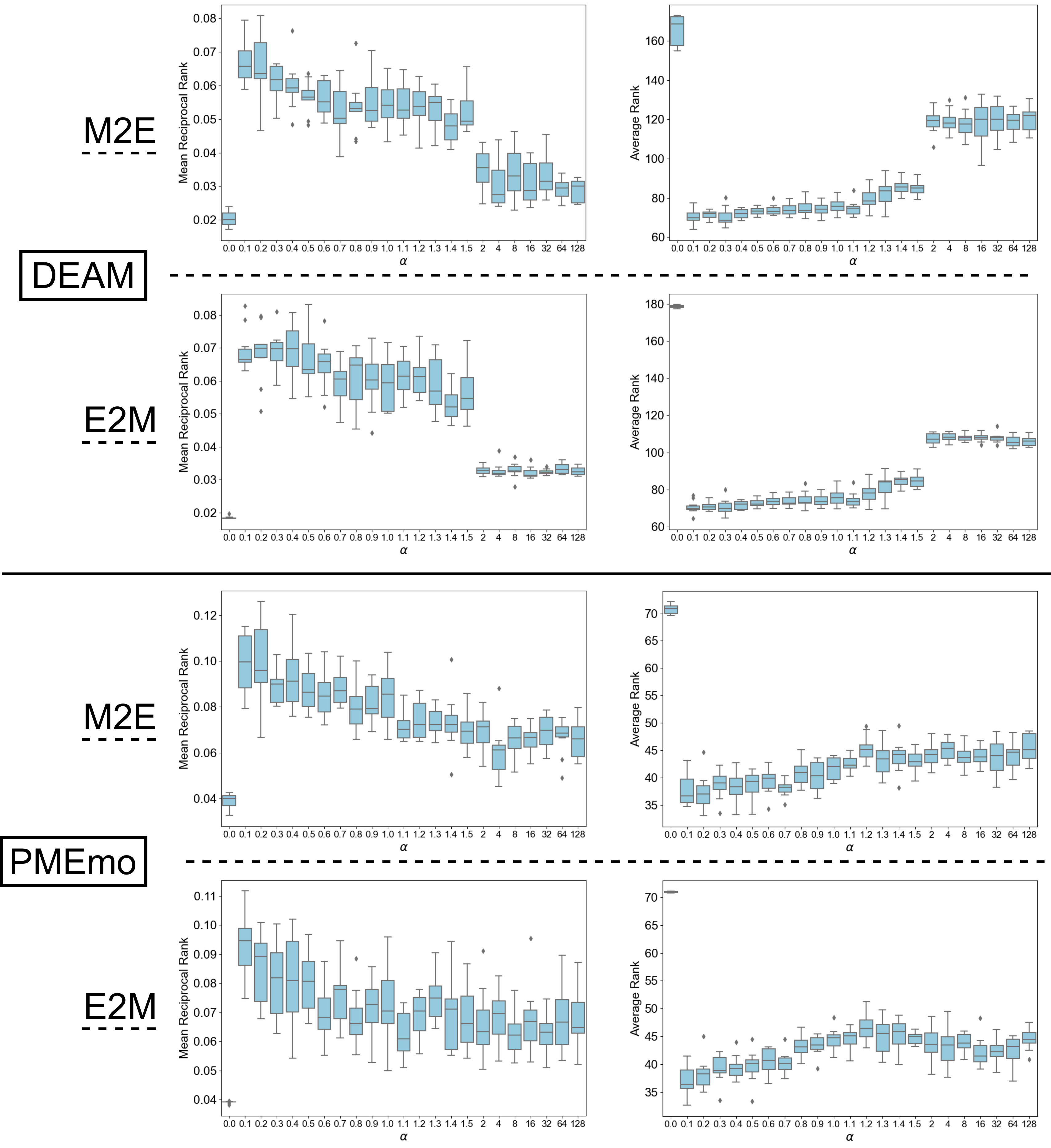}
	\caption{Transition of \textit{Composite-C} performances (MRRs and ARs) obtained using \textit{Cos-Loss} configured by different values of $\alpha$. The optimal $\alpha$ value is selected as $0.3$ and $0.1$ for DEAM and PMEmo, respectively.}
	\label{fig:tuning_alpha_c}
\end{figure*}

The same values as in Section~\ref{sec:tuning_alpha} were tested to tune $\alpha$ for \textit{Cos-Loss}, i.e., values between $0$ and $1.5$ with increments of $0.1$, and powers of $2$ between $2$ and $128$. Using the same box plot format as Figs.~\ref{fig:tuning_P}, \ref{fig:tuning_alpha} and \ref{fig:tuning_lambda}, Fig.~\ref{fig:tuning_alpha_c} displays various performances on DEAM and PMEmo only using \textit{Cos-Loss} configured by different values of $\alpha$. By following the aforementioned criteria addressing the best overall performance on M2E and E2M, $\alpha=0.3$ for DEAM and $\alpha=0.1$ for PMEmo are chosen as the optimal values. The performances of \textit{Cos-Loss} configured with these $\alpha$ values are used in the comparative study in Section~\ref{subsec:eval_composite}.

\begin{figure*}[htbp]
	\centering
	\includegraphics[width=\linewidth]{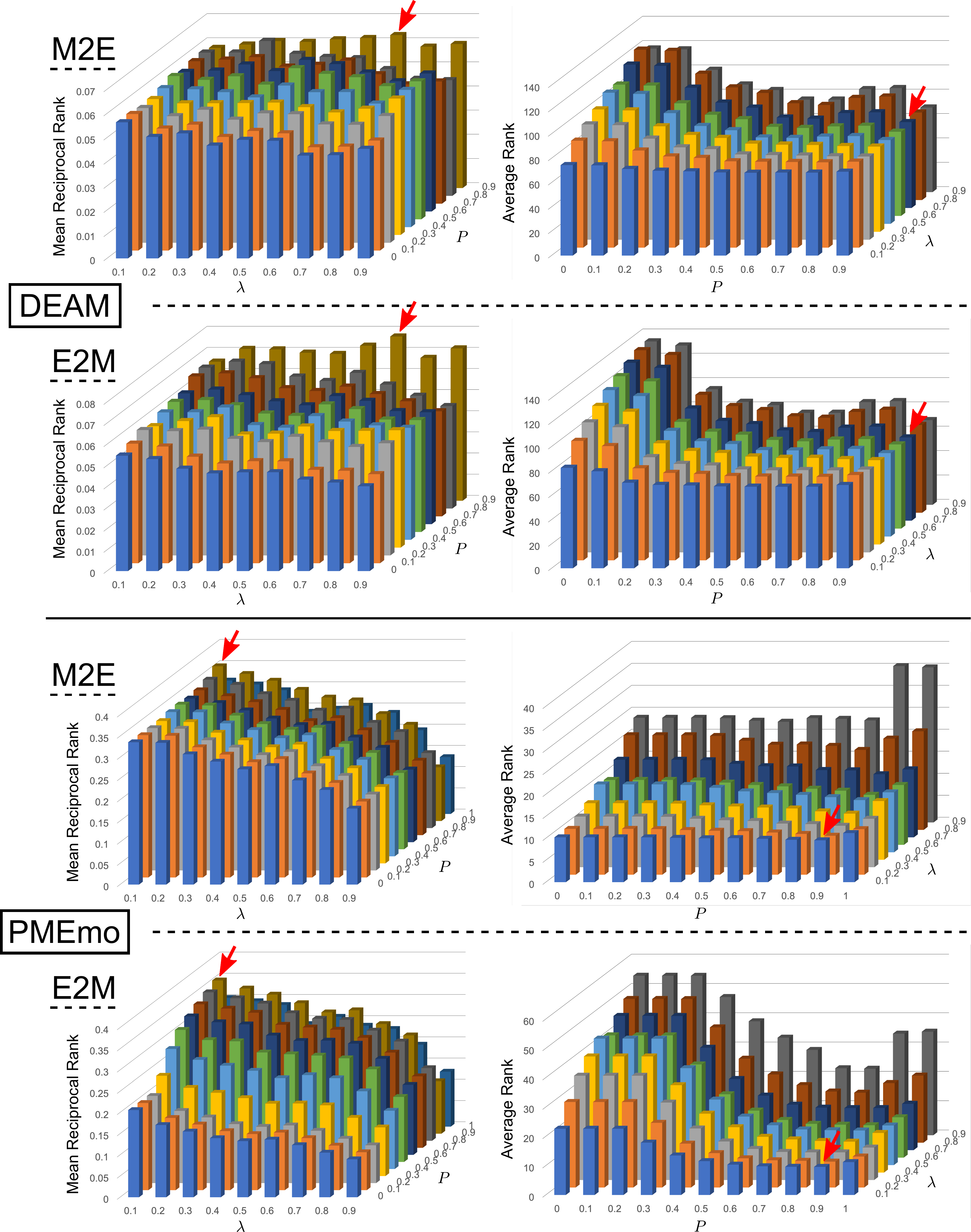}
	\caption{Transitions of \textit{Composite-C} performances (MRRs and ARs) obtained by different pairs of $P$ and $\lambda$ values. Here, $\alpha$ of \textit{Cos-Loss} is set to the optimal values found from Fig.~\ref{fig:tuning_alpha_c}. As indicated by the red arrows, the optimal pair of $P$ and $\lambda$ values is selected as $(P=0.9, \lambda=0.7)$ and $(P=0.9, \lambda=0.1)$ for DEAM and PMEmo, respectively.}
	\label{fig:tuning_grid_P_lambda_c}
\end{figure*}

\subsection{Tuning $P$ and $\lambda$ of \textit{Composite-C} by Grid Search}
\label{sec:tuning_lambda_c}

Fig.~\ref{fig:tuning_grid_P_lambda_c} presents grid search results for different pairs of $P$ and $\lambda$ values, where $P$ ranges between $0$ and the maximum correlation on the considered dataset with an increment of $0.1$ and $\lambda \in \{0.1, \cdots 0.9\}$. Each bar in a three-dimensional bar graph indicates the mean of $10$ performances (i.e., MRRs or ARs) obtained using a pair of $P$ and $\lambda$ values. Like for the hyper-parameter tuning procedure previously described, these $10$ performances are acquired by randomly initialising all parameters in \textit{Composite-C} and randomly splitting a dataset into training and test partitions with a ratio of $8:2$. In addition, the computational cost of grid search can be reduced by considering that $P$ is only related to the test process as described in Section~\ref{sec:tuning_P}. More specifically, after training $10$ \textit{Composite-C} models (i.e., music and emotion encoders, and embedding spaces based on \textit{CCA-Loss} and \textit{Cos-Loss}) using the $\alpha$ value optimised in the previous section and a specific $\lambda$ value, their test processes are repeatedly run to obtain $10$ performances for each of the different $P$ values. Note that to make visual interpretation of the results easier, the axes of $\lambda$ and $P$ are depicted in the horizontal and depth directions for the MRR histograms, while the directions of these axes are swapped for the AR histograms in Fig.~\ref{fig:tuning_grid_P_lambda_c}.

As indicated by the red arrows in Fig.~\ref{fig:tuning_grid_P_lambda_c}, the best overall performance is attained using $P=0.9$ and $\lambda=0.7$ for DEAM and $P=0.9$ and $\lambda=0.1$ for PMEmo. The performances obtained using these optimal $P$ and $\lambda$ are reported as the ones of \textit{Composite-C} in the comparative study in Section~\ref{subsec:eval_composite}. The fact that the \textit{Composite-C} performances are significantly worse than the ones of \textit{Composite} verifies the effectiveness of our proposed EMER-CL based on distribution-based embeddings.

\section{Tuning baselines}
\label{sec:tuning_baselies}

We explain how to tune the hyper-parameters of the regression-based baselines, \textit{RegMLP-M2E}, \textit{RegMLP-E2M}, \textit{RegBiGRU-M2E} and \textit{RegBiGRU-E2M}, used in Section~\ref{subsec:comparison_baseline}. \textit{RegMLP-M2E} and \textit{RegMLP-E2M} employ an MLP, and \textit{RegBiGRU-M2E} and \textit{RegBiGRU-E2M} adopt an RNN based on bidirectional GRU. The inputs of \textit{RegBiGRU-M2E} and \textit{RegMLP-M2E} are a sequence $\boldsymbol{X}^{(m)}$ of $128$-dimensional acoustic features extracted by VGGish, and their mean $\bar{\boldsymbol{x}}^{(m)}$, respectively. The outputs of these baselines are a $2$-dimensional emotion vector $\bar{\boldsymbol{x}}'^{(e)}$ that is a prediction of the average arousal and valence for the music sample corresponding to $\boldsymbol{X}^{(m)}$ or $\bar{\boldsymbol{x}}^{(m)}$. In contrast, \textit{RegBiGRU-E2M} and \textit{RegMLP-E2M} take as input an arousal/valence sequence $\boldsymbol{X}^{(e)}$ of two-dimensional emotion vectors, and their mean $\bar{\boldsymbol{x}}^{(e)}$, respectively. These baselines output a $128$-dimensional acoustic feature $\bar{\boldsymbol{x}}'^{(m)}$ as a prediction of the mean acoustic feature for the music sample associated with $\boldsymbol{X}^{(e)}$ or $\bar{\boldsymbol{x}}^{(e)}$. Below, we describe hyper-parameter tuning for hidden layers used between the above-mentioned inputs and outputs.

Using the same music and emotion encoder architectures as the ones optimised for EMER-CL (as reported in Section~\ref{subsec:impl_details}) did not yield satisfactory performances for the regression-based baselines. More specifically, the M2E baseline obtained an MRR of $0.035 \pm 0.001$ and AR of $109.9 \pm 0.3$ on DEAM, and an MRR of $0.071 \pm 0.004$ and AR of $46.5 \pm 0.7$ on PMEmo. The E2M baseline yielded an MRR of $0.018 \pm 2e^{-6}$ and AR of $181.0 \pm 0.006$ on DEAM, and an MRR of $0.057 \pm 0.003$ and AR of $65.9 \pm 1.2$ on PMEmo. These performances are orders of magnitude worse than the ones of EMER-CL reported in our paper. For this reason, we attempted to improve the performances of the baselines by further tuning their hyper-parameters. 

We focus especially on the number of hidden layers and the number of units per hidden layer, and carry out grid search on them. More specifically, we tested configurations involving a number of layers between one and five, and a number of units per layer in $\{ 16, 32, 64, 128, 256, 512\}$. That is, when using one, two, three, four and five hidden layers, the numbers of possible architectures are $6^1$, $6^2$, $6^3$, $6^4$ and $6^5$, respectively. Thus, our grid search examines in total $9330$ architectures. Note that for \textit{RegBiGRU-M2E} and \textit{RegBiGRU-E2M}, the first hidden layer is defined as a bidirectional GRU layer while the others are defined as FC layers. For \textit{RegMLP-M2E} and \textit{RegMLP-E2M}, all the hidden layers are defined as FC layers. The activation function of units in FC layers is always softplus, and a dropout rate of $0.5$ is applied to FC layers that have $256$ or $512$ units. For a bidirectional GRU layer, the activation functions are used as specified in \cite{Chung_GRU}, and no dropout is employed. Furthermore, the MRR of each architecture is monitored every $100$ epochs until $10000$ epochs, and the model trained at the epoch that yielded the highest MRR is selected for this architecture. One exception is that this performance monitoring is continued until $50000$ epochs unless the training converges after $10000$ epochs. It should be noted that this setting to train the baselines is very favourable, considering that the performance of EMER-CL (i.e., \textit{Composite}) is evaluated only after the last epoch ($5001$ and $10001$ epochs for DEAM and PMEmo, respectively).

Finally, the architecture with the highest MRR after grid search is always used for the comparison with EMER-CL for all baselines, with one exception. In some configurations, an architecture might yield a slightly lower MRR than the highest one, but also a significantly better (lower) AR. It can be considered that the architecture with the highest MRR is relatively subject to overfitting, in which for some queries (i.e., query music samples or query emotions), their associated samples (i.e., emotions or music samples) are ranked at very high positions, while the ranks of samples associated with other queries are very low. In these configurations, we manually select the architecture with the notably better AR among the ones returning the five highest MRRs.

Table~\ref{tab:baseline_param} shows the selected architecture for each of \textit{RegMLP-M2E}, \textit{RegMLP-E2M}, \textit{RegBiGRU-M2E} and  \textit{RegBiGRU-E2M} on DEAM and PMEmo. As seen from Table~\ref{tab:superiority_CMER-CL}, these baselines are significantly outperformed by EMER-CL despite the comprehensive and careful hyper-parameter tuning strategy. This verifies the effectiveness of EMER-CL's embedding approach. 

\begin{table}[!htbp]
	\setlength{\tabcolsep}{5pt}
	\renewcommand{\arraystretch}{1.25}
	\centering
	\caption{Hyper-parameters of the baselines after grid search.}
	\label{tab:baseline_param}
	\vspace{-5pt}
	\begin{tabular}{@{\extracolsep{\fill}}ccc@{\extracolsep{\fill}}}
		\multicolumn{3}{c}{\textbf{DEAM}} \\ \midrule
		Model name & Epochs & Number of units per hidden layer \\ \midrule
		RegMLP-M2E & 8300   & $[128, 64, 16, 128, 512]$ \\
		RegMLP-E2M & 19200  & $[32]$ \\
		RegBiGRU-M2E  & 2000    & $[512, 64, 32, 16, 32]$ \\
		RegBiGRU-E2M  & 9200   & $[128, 128, 32, 128, 128]$ \\ \bottomrule
		\multicolumn{3}{c}{ } \\ 
		\multicolumn{3}{c}{\textbf{PMEmo}} \\ \midrule
		Model name & Epochs & Number of units per hidden layer \\ \midrule
		RegMLP-M2E & 5500   & $[16]$ \\
		RegMLP-E2M & 9900  & $[32, 128, 128]$ \\
		RegBiGRU-M2E  & 7500   & $[512, 32, 32]$ \\
		RegBiGRU-E2M  & 3400   & $[512]$ \\ \bottomrule
	\end{tabular}
\end{table}

\section{Detailed analysis results for M2E}
\label{sec:detailed_m2e}

\begin{figure}[tbp]
	\centering
	\includegraphics[width=\linewidth]{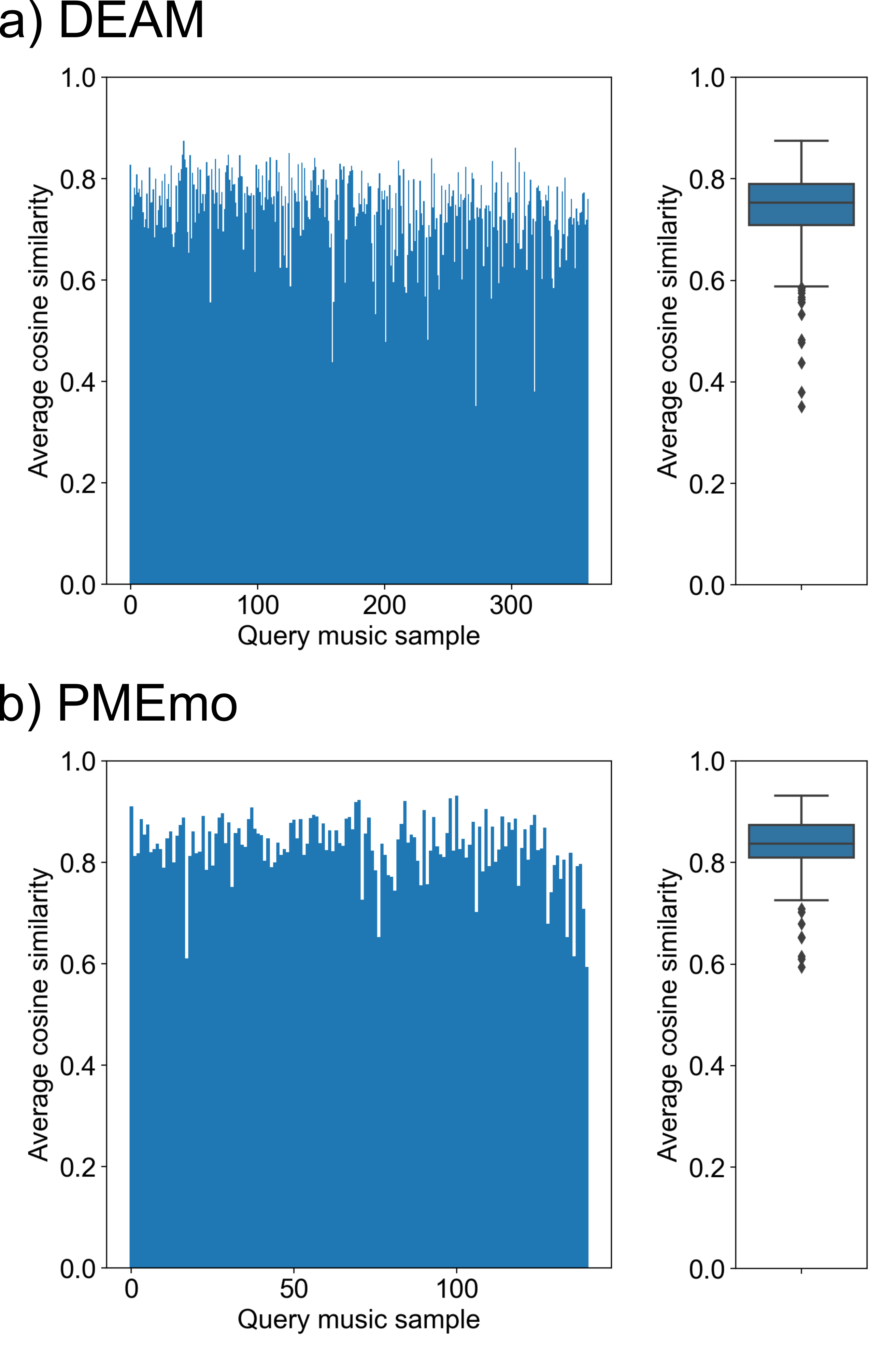}
	\caption{Bar graphs showing average cosine similarities and their box plots for M2E on DEAM and PMEmo.}
	\label{fig:detailed_m2e}
\end{figure}

Fig.~\ref{fig:detailed_m2e} shows the average cosine similarities computed for M2E on DEAM and PMEmo. Cosine similarities are computed between the mean acoustic feature of a query music sample, and the one of the music sample associated to each emotion ranked in the top $5$\% by EMER-CL. A high average cosine similarity means that EMER-CL recognises emotions expressed in music samples that are acoustically similar to a query music sample. In other words, even if EMER-CL fails to properly recognise the ground-truth emotion associated with the query music sample, emotions reasonably relevant to the query are still recognised because they are associated with acoustically similar music samples to the query.

Just like Fig.~\ref{fig:detailed_sim_onebar}, the bar graphs in the left part of Fig.~\ref{fig:detailed_m2e} (a) and (b) are drawn by sorting query music samples in ascending order of the ranks $r_q$ of their associated ground-truth emotions. The more to the left a query music sample is located, the higher its associated emotion is ranked by EMER-CL. It should be noted that the median of pairwise cosine similarities among acoustic features of music samples is $0.674$ and $0.800$ on DEAM and PMEmo, respectively. Each of these numbers can be interpreted as the cosine similarity between the acoustic features of two randomly selected music samples. The bar graphs and box plots in Fig.~\ref{fig:detailed_m2e} show that the median of average cosine similarities is $0.753$ and $0.837$ on DEAM and PMEmo, respectively. These medians are significantly higher than the median of pairwise cosine similarities, which validates the meaningfulness of EMER-CL's M2E results. In addition, as shown by the box plots in Fig.~\ref{fig:detailed_m2e}, the $25$th percentile (first quartile) of average cosine similarities on DEAM and PMEmo are $0.709$ and $0.810$, respectively. The fact that even these $25$th percentiles are higher than the medians of pairwise cosine similarities, indicates that in most cases EMER-CL's M2E works better than randomly selecting an emotion for a query music sample.

\section{About the detailed analysis for E2M on PMEmo}
\label{sec:detailed_pmemo}

\begin{figure}[tbp]
	\centering
	\includegraphics[width=\linewidth]{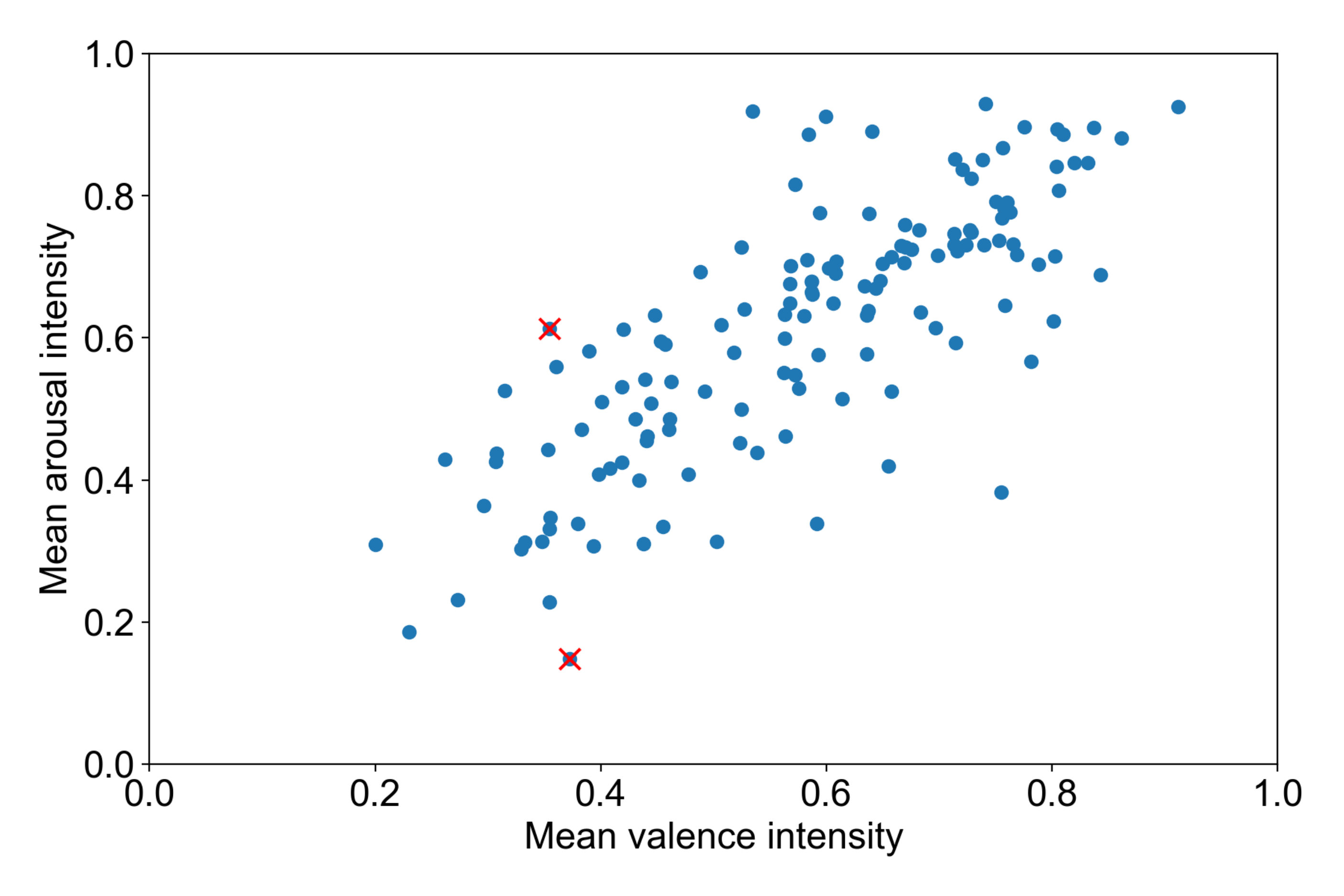}
	\caption{Distribution of PMEmo test emotions in the valence/arousal space.}
	\label{fig:dist_emo_pmemo}
\end{figure}

Fig.~\ref{fig:detailed_sim_onebar} (b) shows that the average cosine similarities in E2M on PMEmo are very high, with a median of 0.994 close to the maximum value of 1. The reasons behind this are discussed below. Fig.~\ref{fig:dist_emo_pmemo} depicts the distribution of $141$ test emotions, each of which is represented by the two-dimensional mean emotion vector $\bar{\boldsymbol{x}}^{(e)}$ of an arousal/valence sequence $\boldsymbol{X}^{(e)}$. Here, arousal and valence intensities in PMEmo are in $[0, 1]$, so all the emotions are distributed only in the first quadrant. Furthermore, the variance of this distribution is small in particular, as illustrated in Fig.~\ref{fig:dist_emo_pmemo}. As a result, the average of pairwise cosine similarities among these emotions is $0.9884 \pm 0.0196$, and even the smallest pairwise cosine similarity characterised by the emotions marked by the crosses in Fig.~\ref{fig:dist_emo_pmemo} is $0.7846$.

It should be noted that if the emotion ratings in Fig.~\ref{fig:dist_emo_pmemo} could be translated so that their origin is $(0,0)$, the range of average cosine similarities would be much wider and a detailed analysis based on them would be much clearer. However, it is difficult to precisely locate the origin of the arousal/valence space in Fig.~\ref{fig:dist_emo_pmemo}. For example, z-normalisation can be carried out so that emotions have zero mean and unit variance, but there is no guarantee that the resulting zero vector corresponds to the origin. Cosine similarities significantly rely on the location of the origin because they measure similarities between the angles of two vectors. In addition, considering Russell’s circumplex model where emotions are circularly located in the arousal/valence space~\cite{CircularModel_Russell}, an angle-based similarity measure like cosine similarity is preferred to other types of measures like Euclidean distance. Thus, reliable analysis is impossible when the exact location of the origin is unknown. For this reason, our detailed analysis on PMEmo is performed using the original arousal and valence intensities without any modification.

\section{Computational Costs of EMER-CL}
\label{sec:comp_costs}

We discuss about the computational costs of EMER-CL by referring to Table~\ref{tab:comp_costs}. Each number in this table indicates the average runtime of $10$ runs to train or test an EMER-CL model with a random initialisation of all parameters and a random $8:2$ split of a dataset into training and test partitions. The runtime for testing the EMER-CL model is the average elapsed time to get an M2E (or E2M) result given a query music sample (or query emotion). The hyper-parameters of the EMER-CL model are set to the optimal values found in Section~\ref{sec:tuning_CMER-CL}. Furthermore, the runtimes in Table~\ref{tab:comp_costs} are measured using a computer equipped with Intel i$9$-$7900$X CPU, $128$GB RAM and NVIDIA RTX $2080$Ti GPU. The codes are written using the TensorFlow library (version $1.15$) based on CUDA version $11.2$\footnote{The codes are available at \url{https://mu-lab.info/naoki_takashima/emer-cl}}.

\begin{table}[!htbp]
	\setlength{\tabcolsep}{10pt}
	\renewcommand{\arraystretch}{1.25}
	\centering
	\caption{Runtimes of EMER-CL.}
	\label{tab:comp_costs}
	\vspace{-5pt}
	\begin{tabular}{@{\extracolsep{\fill}}ccc@{\extracolsep{\fill}}}
		\multicolumn{3}{c}{\textbf{DEAM}} \\ \midrule
		Training & Test (M2E) & Test (E2M) \\ \midrule
		$2118.45$ sec & $0.0046$ sec & $0.0045$ sec \\ \bottomrule
		\multicolumn{3}{c}{ } \\ 
		\multicolumn{3}{c}{\textbf{PMEmo}} \\ \midrule
		Training & Test (M2E) & Test (E2M) \\ \midrule
		$9472.44$ sec & $0.0025$ sec & $0.0019$ sec \\ \bottomrule
	\end{tabular}
\end{table}

As it can be seen from the first column in Table~\ref{tab:comp_costs}, training an EMER-CL model on PMEmo takes significantly longer time than training it on DEAM, even though the training partition of PMEmo only contains $561$ music-emotion pairs. One main reason is that both of the music and emotion encoders for PMEmo are defined as RNNs that need to process the acoustic feature or emotion vector sequentially in time. To improve the scalability of EMER-CL, we plan to define the music and emotion encoders as self-attention models that can perform batch processing of acoustic features or emotion vectors at all times~\cite{attention_is_all_you_need}.

The second and last columns in Table~\ref{tab:comp_costs} show the very short runtimes of EMER-CL's test process. This is due to the fact that the test partitions of DEAM and PMEmo only contain $361$ and $141$ samples (i.e., music samples for E2M or emotions for M2E), respectively. Nevertheless, since the runtime of EMER-CL's test process scales linearly with the number of samples, the test process would still be expected to finish within seconds even with a number of samples three orders of magnitude higher than the currently tested numbers.

\section{t-SNE plots of embeddings on the PMEmo dataset}
\label{sec:tsne-pmemo}
In a similar way as for the DEAM dataset, we plotted the t-SNE projections of the correlation-based and probabilistic-based embeddings produced by both the music ($\boldsymbol{\phi}^{(m)}$, $\boldsymbol{\mu}^{(m)}$ and $\boldsymbol{\sigma}^{(m)}$) and emotion models ($\boldsymbol{\phi}^{(e)}$, $\boldsymbol{\mu}^{(e)}$ and $\boldsymbol{\sigma}^{(e)}$) trained on the PMEmo dataset. Plots were obtained for examples in both the training and testing sets. For each example of either dataset, we first computed the average emotion vector over time $\bar{\boldsymbol{x}}^{(e)}_{j}$ and used it to label the t-SNE projections in terms of which quadrant in the arousal/valence space (HA/HV, HA/LV, LA/LV or LA/HV) each music sample $\boldsymbol{X}^{(m)}_j$ or mean emotion sequence $\boldsymbol{x}^{(e)}_j$ was associated with.

\begin{figure*}[tbp]
	\centering
	\includegraphics[width=\linewidth]{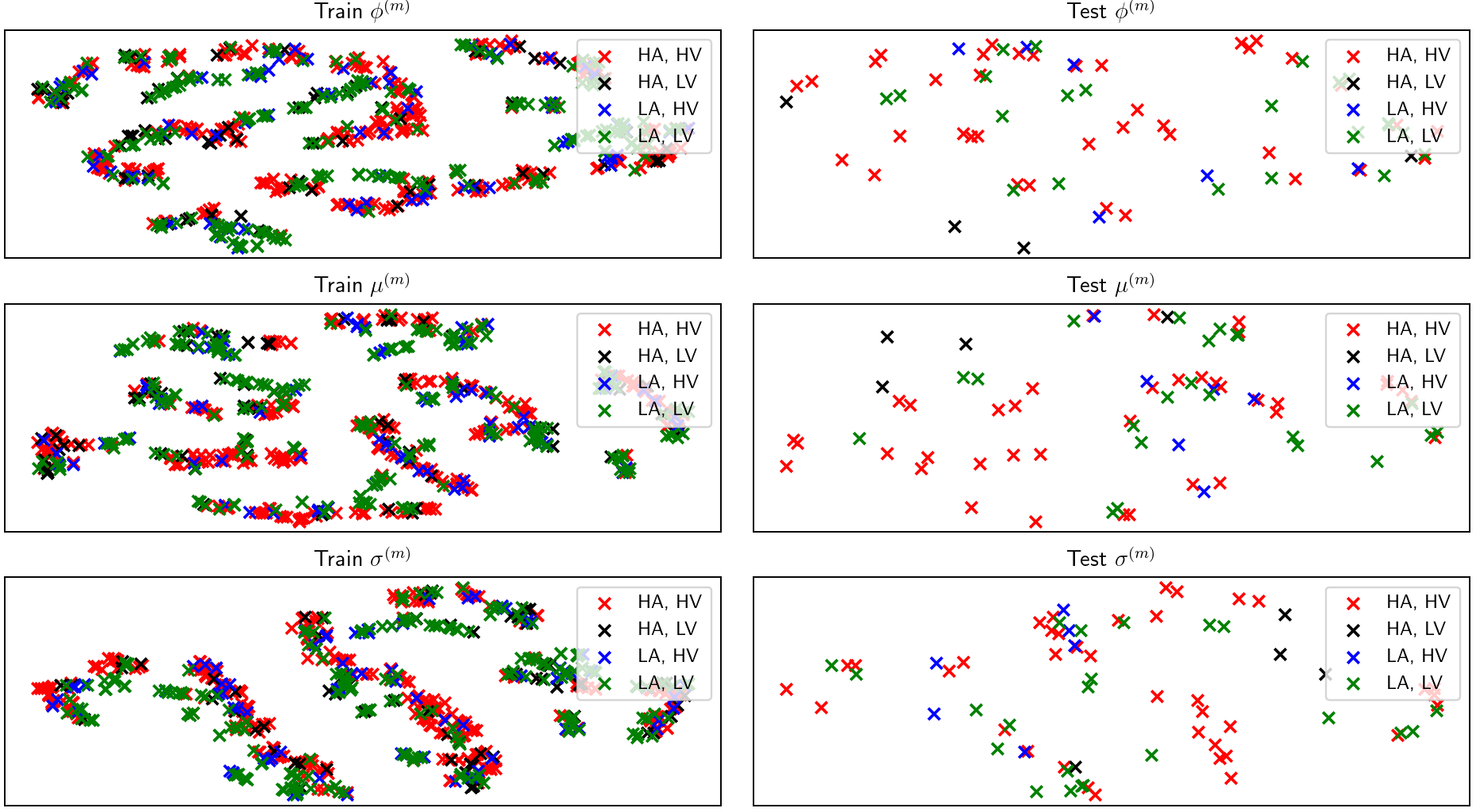}
	\caption{t-SNE projections of the music embeddings $\phi^{(m)}$, $\mu^{(m)}$ and $\sigma^{(m)}$ (respectively first, second and third rows) on the PMEmo training (left column) and testing (right column) sets. The t-SNE projections are labelled with their associated emotional quadrants, i.e., high arousal/high valence (HA, HV), high arousal/low valence (HA, LV), low arousal/low valence (LA, LV) and low arousal/high valence (LA, HV).}
	\label{fig:tsne-pmemo-music}
	% \vspace{-0.5cm}
\end{figure*}

\begin{figure*}[tbp]
	\centering
	\includegraphics[width=\linewidth]{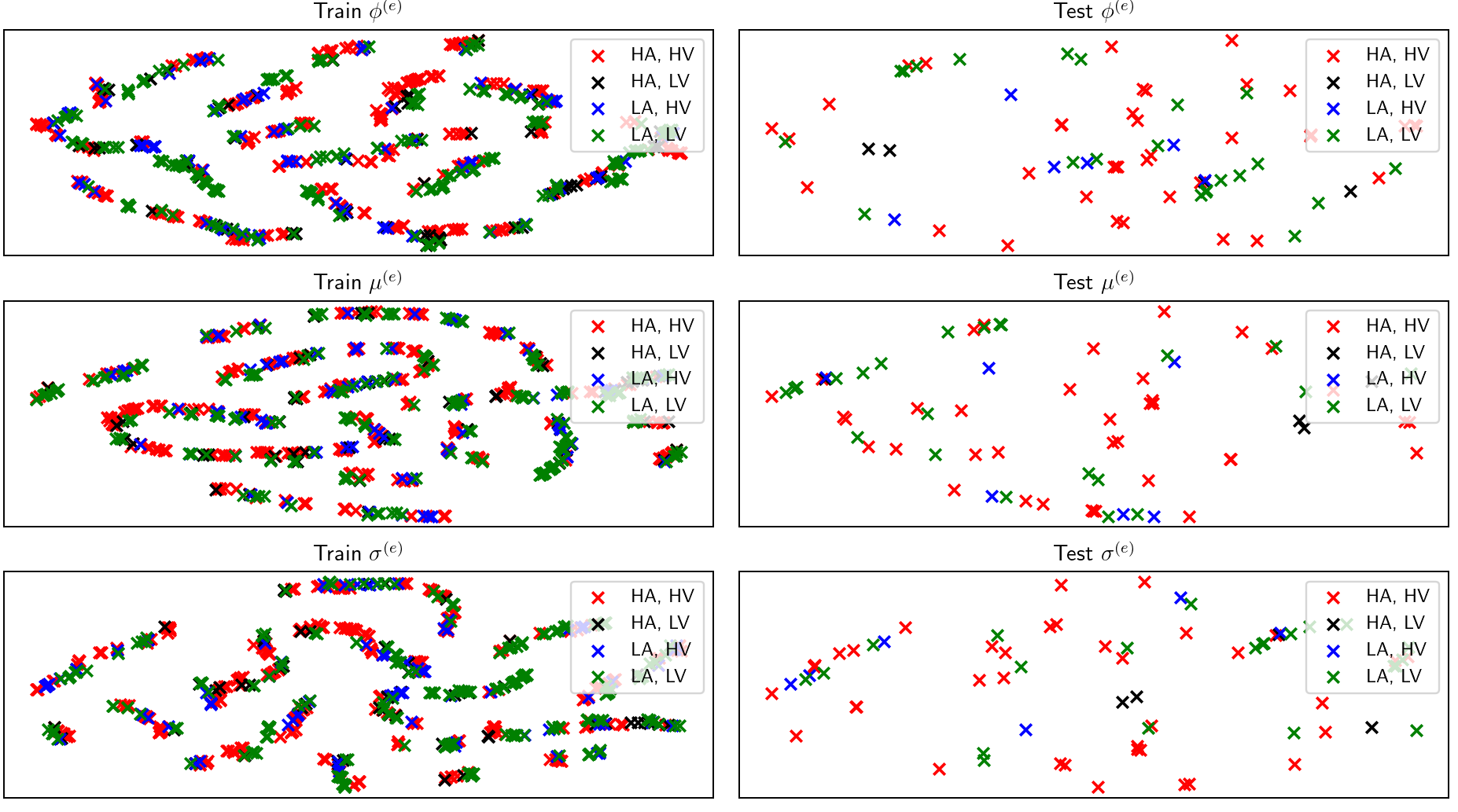}
	\caption{t-SNE projections of the emotion embeddings $\phi^{(e)}$, $\mu^{(e)}$ and $\sigma^{(e)}$ (respectively first, second and third rows) on the PMEmo training (left column) and testing (right column) sets. The t-SNE projections are labelled with their associated emotional quadrants, i.e., high arousal/high valence (HA, HV), high arousal/low valence (HA, LV), low arousal/low valence (LA, LV) and low arousal/high valence (LA, HV).}
	\label{fig:tsne-pmemo-emotion}
	% \vspace{-0.5cm}
\end{figure*}

Unlike for the DEAM dataset, the definition of the emotional quadrants on PMEmo is not trivial. This is due to the fact that both arousal and valence ratings are projected into the $[0,1]$ interval, and that the dataset includes exclusively pop songs which skew it strongly towards the quadrant high arousal/high valence (HA/HV). Taking the centre of the emotional space $(0.5, 0.5)$ as cut-off point for the definition of the quadrants leads to a very imbalanced data repartition with $476$, $53$, $106$ and $66$ songs out of $701$ associated to the HA/HV, HA/LV, LA/LV and LA/HV quadrants, respectively. To mitigate the effects of this imbalance, we instead take the barycentre of all $701$ PMEmo music samples in the arousal/valence space $(0.61, 0.63)$ as mid-point to define the quadrants. This leads to a more balanced repartition of $306$, $77$, $84$ and $234$ songs associated to the HA/HV, HA/LV, LA/LV and LA/HV quadrants, respectively. The t-SNE plots of embeddings obtained with these emotion annotations are provided in Figs.~\ref{fig:tsne-pmemo-music} and~\ref{fig:tsne-pmemo-emotion} for the music and emotion embeddings, respectively.

These figures show that embeddings tend to be grouped by their associated emotional quadrants on both the training and testing sets, although this trend on the PMEmo dataset is less clear than the one for the DEAM dataset, as some groups mixing embeddings associated with different quadrants can also be seen. We hypothesise that this is due to the fact that the PMEmo dataset aggregates music samples belong to a single genre (pop songs) that are quite similar to each other in terms of audio features and elicited emotions. Despite this, music samples or emotions associated to the same quadrants still tend to be located in the same neighbourhoods of the embedding space.

\bibliographystyle{unsrt}
\bibliography{refs}

% \begin{thebibliography}{00}
% 
% \bibitem{b1} G. O. Young, ``Synthetic structure of industrial plastics,'' in \emph{Plastics,} 2\textsuperscript{nd} ed., vol. 3, J. Peters, Ed. New York, NY, USA: McGraw-Hill, 1964, pp. 15--64.
% 
% \end{thebibliography}

\begin{IEEEbiography}[{\includegraphics[width=1in,height=1.25in,clip,keepaspectratio]{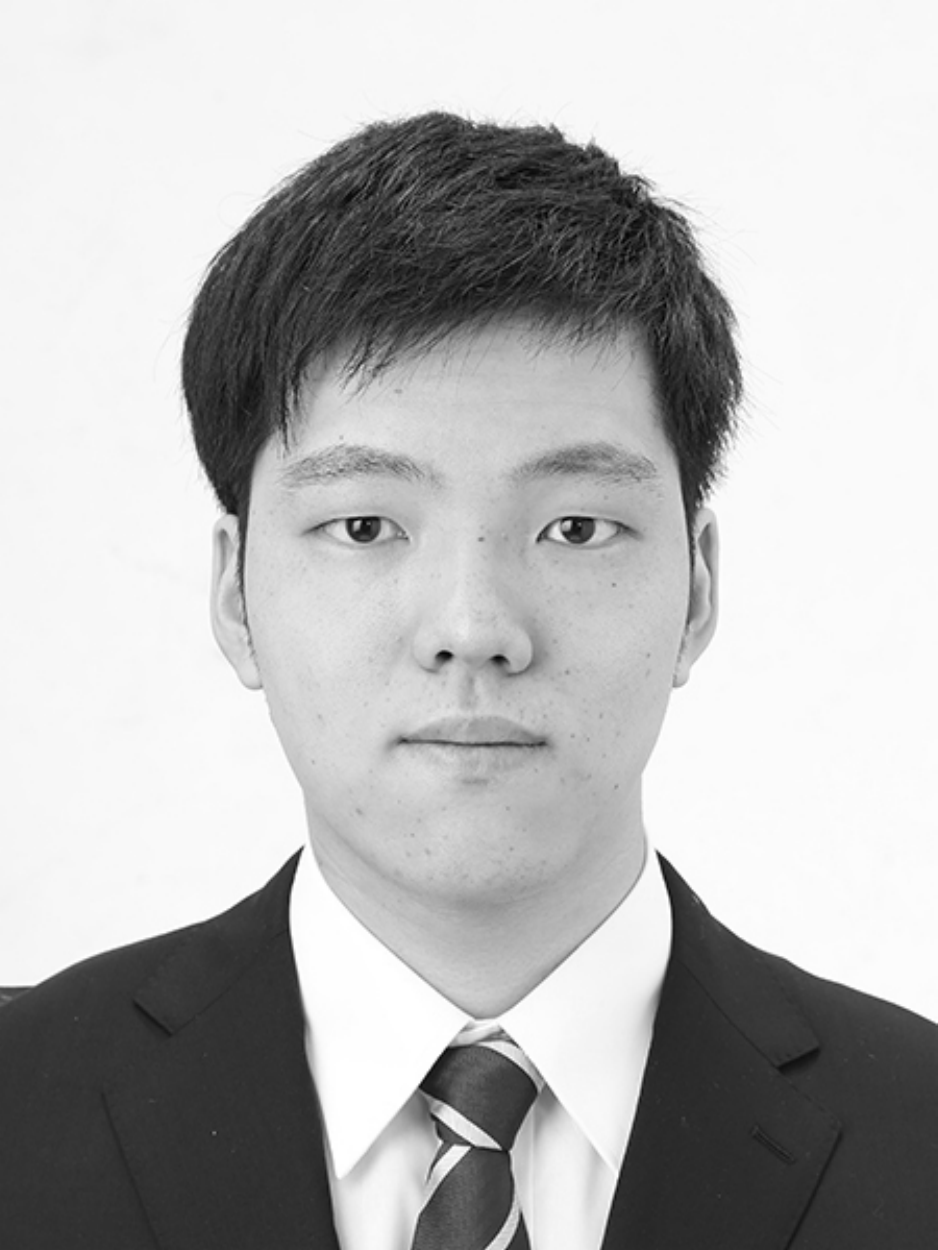}}]{Naoki Takashima} received his B.E. and M.E. degrees in engineering from Kindai University, Osaka Japan, in 2020 and 2022, respectively. Almost all of the work needed to prepare this paper was done when he was pursuing his M.E. degree. He is now working at Speee Inc. His research interests includes affective computing, music emotion recognition and deep learning.
\end{IEEEbiography}

\begin{IEEEbiography}[{\includegraphics[width=1in,height=1.25in,clip,keepaspectratio]{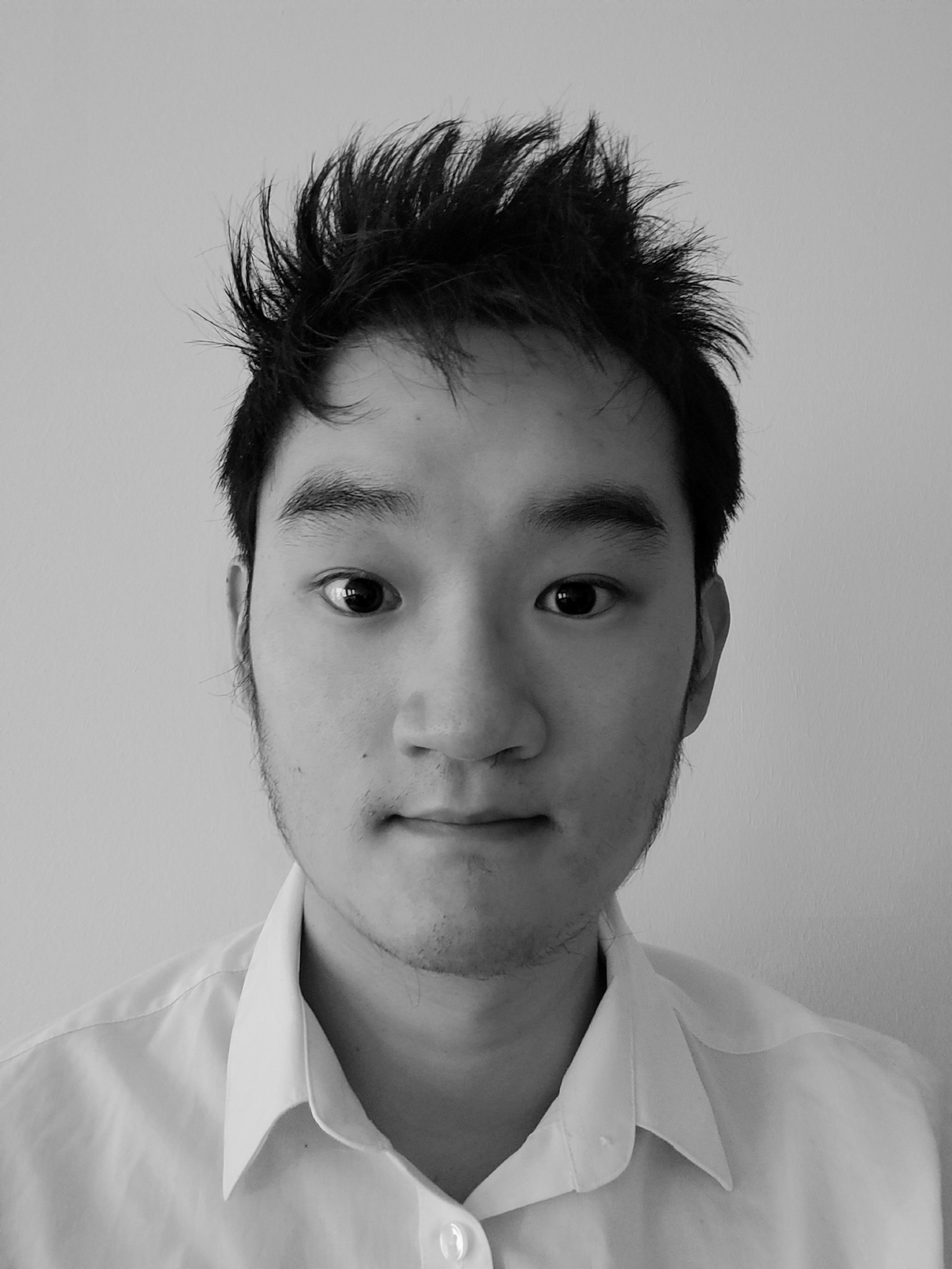}}]{Frédéric~Li} received his engineering degree (equivalent of a M. Sc. degree) in the French Grande Ecole ENSTA Paristech with a specialisation in Robotics and Embedded Systems in 2015. He worked as a research assistant under the supervision of Prof. Dr.-Ing. habil. M. Grzegorzek in the Pattern Recognition Group of the University of Siegen (Germany) between 2016 and 2019, and since 2019 in the Institute of Medical Informatics of the University of Lübeck (Germany), where he obtained his doctoral degree in 2021. His research interests include ubiquitous computing, time-series classification, feature extraction, deep learning, and transfer learning.
\end{IEEEbiography}

\begin{IEEEbiography}[{\includegraphics[width=1in,height=1.25in,clip,keepaspectratio]{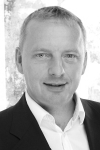}}]{Marcin Grzegorzek} is full professor of medical informatics at the University of Lübeck leading the Medical Data Science Lab. He obtained his master degree in computer science from the Silesian University of Technology in Gliwice in 2002, his doctor of engineering degree with distinction from the University of Erlangen-Nürnberg in 2007 and his habilitation degree from the AGH University of Science and Technology in Kraków in 2014. Marcin's research areas include pattern recognition, machine learning, data science and sensor data analysis for health-related applications. He and his team conceptualise, implement and evaluate new algorithms for automated analysis of human-related data (e.g., wearable sensor data) and demonstrate their applicability in real-world scenarios. Prof. Grzegorzek has published more than 100 scientific peer-reviewed articles, has been project leader in more than 10 third-party funded research projects and acted as supervisor in 9 completed doctoral procedures.
\end{IEEEbiography}

\begin{IEEEbiography}[{\includegraphics[width=1in,height=1.25in,clip,keepaspectratio]{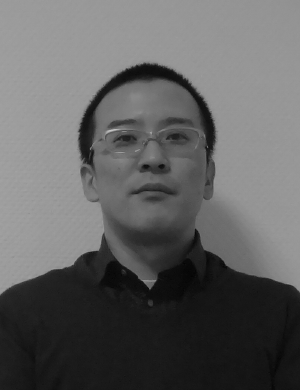}}]{Kimiaki~Shirahama} received his B.E., M.E. and D.E. degrees in engineering from Kobe University, Japan in 2003, 2005 and 2011, respectively. After working as an assistant professor at Muroran Institute of Technology, Japan, he worked as a postdoctoral researcher at Pattern Recognition Group in University of Siegen, Germany from 2013 to 2018. He then worked as an associate professor at Kindai University, Japan from 2018 to 2023. He is currently working as an associate professor at Doshisha University, Japan. His research interests include multimedia data processing, machine learning, data mining and sensor-based human activity recognition. He is a member of ACM SIGKDD, ACM SIGMM, the Institute of Image Information and Television Engineers in Japan (ITE), Information Processing Society of Japan (IPSJ) and the Institute of Electronics, Information and Communication Engineering in Japan (IEICE).
\end{IEEEbiography}

\EOD

\end{document}